\documentclass[prb,aps,twocolumn,superscriptaddress,longbibliography]{revtex4-2}

\usepackage{amsmath,amssymb}
\usepackage{graphicx}
\usepackage{bm}
\usepackage{hyperref}

\newcommand{\lno}{La$_3$Ni$_2$O$_7$}
\graphicspath{{figures/}}
\begin{document}

\title{Strain-tuned structural, electronic, and superconducting \\ properties of thin-film La$_3$Ni$_2$O$_7$}

\author{Sreekar Bheemavarapu}

\date{\today}

\begin{abstract}
The recent discovery of high-temperature superconductivity in \lno\ under ambient-pressure in strained thin films raises the question of how superconductivity can be optimized through strain. In this work, we investigate the strain-dependent electronic structure and superconducting transition temperature ($T_c$) of \lno\ using density functional theory combined with random phase approximation spin-fluctuation calculations. We find that biaxial strain acts as a tuning parameter for Fermi surface topology and magnetic correlations. Large tensile strain drives a Lifshitz transition characterized by a $d_{z^2}$ band crossing, leading to a sharp increase in the density of states and theoretical pairing strength. However, this is accompanied by a large increase in magnetic proximity, suggesting strong competition with spin-density-wave order. Conversely, under compressive strain, we identify a structurally selective $T_c$ enhancement restricted to the high-symmetry $I4/mmm$ phase. This effect is driven by the straightening of Ni--O--Ni bonds and the emergence of a $\Gamma$-centered hole pocket, yielding $T_c$ values consistent with recent thin-film experiments. Our results highlight the balance between structural symmetry, electronic topology, and magnetic instability in nickelates, and provides a theoretical framework for optimizing superconductivity via strain engineering.
\end{abstract}

\maketitle

\section{Introduction}
Nickelates have long been explored as potential analogs to cuprate high-$T_c$ superconductors~\cite{anisimovElectronicStructurePossible1999,leeInfinitelayer$mathrmLamathrmNimathrmO_2$$mathrmNi^1+$2004,liSuperconductivityInfinitelayerNickelate2019,sunSignaturesSuperconductivity802023}. While superconductivity was demonstrated in infinite-layer nickelates ($R$NiO$_2$), critical temperatures were modest ($T_c \sim 9\text{--}15$ K) and synthesis proved challenging~\cite{liSuperconductivityInfinitelayerNickelate2019}. Recently, the bilayer Ruddlesden-Popper nickelate La$_3$Ni$_2$O$_7$ has attracted attention for superconducting signatures near 80 K under high hydrostatic pressure~\cite{sunSignaturesSuperconductivity802023, wangPressureInducedSuperconductivityPolycrystalline2024, houEmergenceHighTemperatureSuperconducting2023}.

However, the bulk high-$T_c$ state requires high hydrostatic pressure ($\gtrsim 14$ GPa) to suppress a competing density-wave order~\cite{sunSignaturesSuperconductivity802023, wangPressureInducedSuperconductivityPolycrystalline2024, zhangHightemperatureSuperconductivityZero2024, liuEvidenceChargeSpin2022, chenEvidenceSpinDensity2024}. This requirement limits applications and complicates experimental probes like ARPES or STM. Thus, the recent discovery that epitaxial strain in thin films can stabilize ambient-pressure superconductivity with onset $T_c$ between 26-42 K has led to much interest on the structural, electronic, and superconducting properties of thin film \lno~\cite{koSignaturesAmbientPressure2025, zhouAmbientpressureSuperconductivityOnset2025,bhattaStructuralElectronicEvolution2025,osadaStraintuningSuperconductivityLa3Ni2O72025,bhattResolvingStructuralOrigins2025}.

Strain mimics some effects of hydrostatic pressure, but introduces distinct interface and symmetry effects~\cite{koSignaturesAmbientPressure2025, zhouAmbientpressureSuperconductivityOnset2025, geislerFermiSurfaceReconstruction2025, rhodesStructuralRoutesStabilize2024}. In addition, unlike the nominal $3d^9$ configuration of infinite-layer nickelates, La$_3$Ni$_2$O$_7$ possesses a $3d^{7.5}$ mixed-valence state, with its physics defined by the combination of the$d_{x^2-y^2}$ orbital (intralayer hopping) and the $d_{z^2}$ orbital, which forms strong interlayer bands via the apical oxygen~\cite{luoBilayerTwoOrbitalModel2023, christianssonCorrelatedElectronicStructure2023, lechermannElectronicCorrelationsSuperconducting2023, yangOrbitaldependentElectronCorrelation2024}. These traits and others distinguish thin film \lno from its previously studied variants.

In the bulk, superconductivity coincides with a structural transition from orthorhombic $Amam$ to higher-symmetry $Fmmm$ or $I4/mmm$ phases, although samples often contain mixed phases~\cite{puphalUnconventionalCrystalStructure2024, wangStructureResponsibleSuperconducting2024, zhangStructuralPhaseTransition2024}. Straightening the Ni-O-Ni bond angles, particularly along the $c$-axis, is believed to maximize the superexchange interaction $J_{\perp}$~\cite{xieStrongInterlayerMagnetic2024, luInterlayerCouplingDrivenHighTemperatureSuperconductivity2024}. Oxygen nonstoichiometry and intergrowths further complicate interpretation~\cite{koSignaturesAmbientPressure2025, puphalUnconventionalCrystalStructure2024}. Evidence suggests superconductivity is mediated by spin fluctuations~\cite{chenElectronicMagneticExcitations2024}, with strong Hund's coupling locking orbital spins to facilitate interlayer pairing~\cite{liu$s^ifmmodepmelsetextpmfi$WavePairingDestructive2023,ohTypeII$tensuremathJ$Model2023}.

The necessity of the $d_{z^2}$-derived hole pocket ($\gamma$ pocket) crossing the Fermi level for superconductivity remains debated~\cite{shaoBandStructurePairing2025, jiangHighTemperatureSuperconductivity2024}. While some models suggest this metallization is essential, recent thin-film works report superconductivity even when this band remains occupied~\cite{geislerFermiSurfaceReconstruction2025,bhattaStructuralElectronicEvolution2025}. Competition between $d$-wave and $s_\pm$-wave pairing symmetries is also unresolved~\cite{guEffectiveModelPairing2025}. While $s_\pm$ is favored by strong interlayer coupling ($J_{\perp}$), strain-tunable crystal field splitting may drive a transition to $d$-wave or $d+is$ states~\cite{sakakibaraPossibleHigh$T_c$2024, quBilayer$ttextensuremathJtextensuremathJ_ensuremathperp$Model2024, xiaSensitiveDependencePairing2025}. 

In this study, we use Density Functional Theory (DFT) and Wannier90~\cite{giannozziQUANTUMESPRESSOModular2009, mostofiWannier90ToolObtaining2008} to construct multi-orbital tight binding models, which are then used with the Random Phase Approximation (RPA) to solve the linearized Eliashberg equation and calculate the pairing eigenvalue $\lambda$, eventually determining quantitative $T_c$ estimates~\cite{graserNeardegeneracySeveralPairing2009}. We map out the structural and electronic evolution of \lno under three phases ($Amam$, $Amam$ with constrained in-plane lattice parameters, and $I4/mmm$) across a range of biaxial strains ($-2.5\%$ to $+2.0\%$). Our results reveal the interplay between strain and the structural symmetry, electronic topology, and magnetic fluctuations that govern superconductivity in \lno~\cite{lechermannElectronicCorrelationsSuperconducting2023, chenElectronicMagneticExcitations2024, luInterlayerCouplingDrivenHighTemperatureSuperconductivity2024}.
\section{Computational Methods}

\subsection{DFT Structural Relaxation}
We studied the structural response of \lno to biaxial strain for three phases: $Amam$, $Amam$ with constrained in-plane lattice parameters ($a = b$), and $I4/mmm$. All calculations used a 24-atom simulation cell. The $Amam$ phases employed a rotated $\sqrt{2} \times \sqrt{2}$ supercell, while the $I4/mmm$ phase used a conventional cell.

The unstrained $Amam$ reference was obtained by full variable-cell relaxation of the experimental structure~\cite{lingNeutronDiffractionStudy2000}. The $Amam$ $(a = b)$ reference enforced equal in-plane parameters ($a'=b'=\sqrt{ab}$), relaxing only the $c$-axis and internal coordinates. The $I4/mmm$ reference was initialized from experiment~\cite{vaulxPressureStrainEffects2025} with in-plane parameters matched to the $Amam$ $(a = b)$ reference ($a_{I4} = a_{Amam}/\sqrt{2}$), followed by relaxation of the $c$-axis and internal coordinates.

Biaxial strain was applied by fixing in-plane parameters ($a = (1 + \epsilon)a_0$, $b = (1 + \epsilon)b_0$) and relaxing the $c$-axis and internal coordinates. We considered strains $\epsilon \in \{-2.5\%, -2.0\%, -1.5\%, -1.0\%, -0.5\%, 0\%, +1.0\%, +2.0\%\}$.

\subsection{Electronic Structure Calculations}
DFT calculations were performed using \textsc{Quantum ESPRESSO}~\cite{giannozziQUANTUMESPRESSOModular2009, giannozziAdvancedCapabilitiesMaterials2017} in the nonmagnetic state. We employed the DFT+$U$ formalism~\cite{anisimovBand1991, dudarevElectronenergylossSpectraStructural1998} with $U_{\text{Ni}-d} = 3.5$~eV, consistent with prior studies~\cite{bhattaStructuralElectronicEvolution2025,yangOrbitaldependentElectronCorrelation2024}. We used GBRV PBE pseudopotentials~\cite{garrityPseudopotentialsHighthroughputDFT2014, perdewGeneralizedGradientApproximation1996} with a 40~Ry kinetic energy cutoff and 320~Ry charge density cutoff. Occupations were treated with Marzari-Vanderbilt cold smearing~\cite{marzariThermalContractionDisordering1999} (width 0.02~Ry). The energy convergence threshold was $10^{-8}$~Ry.

Relaxations used a $10\times10\times5$ k-mesh for $Amam$ phases and $12\times12\times4$ for $I4/mmm$. Self-consistent calculations used $12\times12\times6$ ($Amam$) and $17\times17\times4$ ($I4/mmm$) meshes to maintain comparable sampling density. Non-self-consistent calculations used $16\times16\times8$ ($Amam$) and $22\times22\times4$ ($I4/mmm$) meshes.

\subsection{Wannierization}
Maximally localized Wannier functions were constructed via \textsc{Wannier90}~\cite{mostofiWannier90ToolObtaining2008, pizziWannier90CommunityCode2020} using a minimal two-orbital Ni $e_g$ model ($d_{z^2}$, $d_{x^2-y^2}$)~\cite{luoBilayerTwoOrbitalModel2023}. We used a disentanglement window of ($E_F - 2$~eV, $E_F + 3.5$~eV) and a frozen window of ($E_F - 1.0$~eV, $E_F + 1.0$~eV) across all strains and phases.

\subsection{Susceptibility and RPA Formalism}
Multi-orbital susceptibility was calculated from the Wannier models. The bare susceptibility $\chi_0^{pqrs}(\mathbf{q})$ was computed on a $64 \times 64$ $\mathbf{q}$-mesh using a $64 \times 64$ $\mathbf{k}$-mesh and broadening $\eta = 2$ meV, including temperature dependence via Fermi-Dirac distributions:
\begin{equation}
\begin{split}
\chi_0^{pqrs}(\mathbf{q}) = -\frac{1}{N_k} \sum_{\mathbf{k}, \mu, \nu} \frac{f(E_\nu(\mathbf{k}+\mathbf{q})) - f(E_\mu(\mathbf{k}))}{E_\nu(\mathbf{k}+\mathbf{q}) - E_\mu(\mathbf{k}) + i\eta} \\
\times U_{\mu}^p(\mathbf{k})^* U_{\mu}^q(\mathbf{k}) U_{\nu}^r(\mathbf{k}+\mathbf{q})^* U_{\nu}^s(\mathbf{k}+\mathbf{q})
\end{split}
\end{equation}
where $E_\mu(\mathbf{k})$ and $U_{\mu}^p(\mathbf{k})$ are eigenvalues and eigenvectors, and $f(E)$ is the Fermi-Dirac distribution.

Spin ($\chi_s$) and charge ($\chi_c$) susceptibilities were obtained in the random phase approximation (RPA):
\begin{align}
\chi_s^{RPA}(\mathbf{q}) &= \chi_0(\mathbf{q}) [1 - U_s \chi_0(\mathbf{q})]^{-1} \\
\chi_c^{RPA}(\mathbf{q}) &= \chi_0(\mathbf{q}) [1 + U_c \chi_0(\mathbf{q})]^{-1}
\end{align}
where $U_s$ and $U_c$ are orbital-space interaction matrices including intra-orbital $U$, inter-orbital $U'$, Hund's coupling $J$, and pair hopping $J'$ ($U' = U - 2J$, $J' = J$). Note that this $U$ differs from the DFT+$U$ one, being a low-energy effective interaction.
We determined the interaction strength $U$ using a Stoner criterion: for each structure, $U$ was set to $0.99 U_{crit}$, where $U_{crit}$ is the critical value for magnetic instability (i.e. the $U$ that makes the largest eigenvalue of $U_s \chi_0(\mathbf{q})$ equal 1), with fixed $J/U = 0.15$.

The singlet pairing vertex $\Gamma(\mathbf{k}, \mathbf{k}')$ was constructed from spin and charge fluctuations:
\begin{equation}
\Gamma(\mathbf{q}) = \frac{3}{2} U_s \chi_s^{RPA}(\mathbf{q}) U_s - \frac{1}{2} U_c \chi_c^{RPA}(\mathbf{q}) U_c + \frac{1}{2} (U_s + U_c)
\end{equation}
where $\mathbf{q} = \mathbf{k} - \mathbf{k}'$. We also included the symmetric $\Gamma(\mathbf{k}+\mathbf{k}')$ contribution.
We maintained a fixed filling of $n = 1.5$ electrons per Ni site across all strains and phases by adjusting the chemical potential.

\subsection{\texorpdfstring{$T_c$ Estimation}{Tc Estimation}}
$T_c$ was determined by solving the linearized Eliashberg gap equation. The Fermi surface was discretized using k-points within $|E_n(\mathbf{k}) - \mu| < 5$ meV, subsampled to 600-1600 points if necessary. Integration weights $w_i = \Delta l_i / (2\pi v_F(\mathbf{k}_i))$ were assigned using an arc-length method. The gap equation $\lambda \Delta(\mathbf{k}) = - \sum_{\mathbf{k}'} \Gamma(\mathbf{k}, \mathbf{k}') w_{\mathbf{k}'} \Delta(\mathbf{k}')$ was solved by symmetrizing the kernel $\tilde{\Gamma}_{ij} = \sqrt{w_i} \Gamma_{ij} \sqrt{w_j}$. We included a logarithmic cutoff factor $\ln(1.13 \omega_c / T)$ with $\omega_c = 60$ meV. $T_c$ is defined as the temperature where the leading eigenvalue $\lambda = 1$.

\section{Results}

\subsection{Structural energetics under biaxial strain}
\begin{figure}[t]
\centering  
\includegraphics[width=0.8\columnwidth]{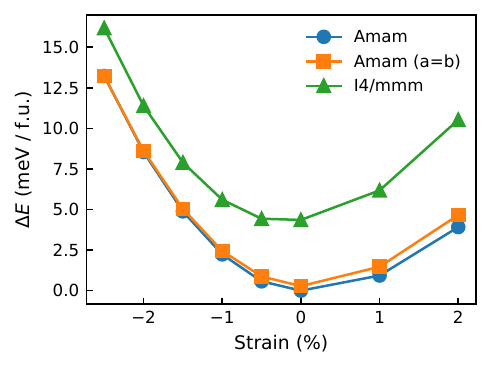}
\caption{Total energy as a function of biaxial strain for each phase, referenced to the minimum-energy structure (unstrained $Amam$). Energy differences remain on the order of 10 meV/f.u. across the strain range shown.}
\label{fig:energy}
\end{figure}

Figure~\ref{fig:energy} shows the total energy vs. biaxial strain for each phase, referenced to the lowest energy strain/phase (unstrained $Amam$). Energy differences are $\sim$10 meV/f.u. The $Amam$ and $Amam$ ($a = b$) curves are nearly identical, indicating a low energy cost for enforcing in-plane tetragonality. The $I4/mmm$ phase remains higher in energy throughout, but becomes competitive with the $Amam$ phases under large compressive strain. On the other hand, tensile strain stabilizes the lower-symmetry phases, increasing the energy difference between the $Amam$ phases and $I4/mmm$.

\subsection{Structural response to biaxial strain}
\begin{figure}[t]
\centering
\includegraphics[width=0.8\columnwidth]{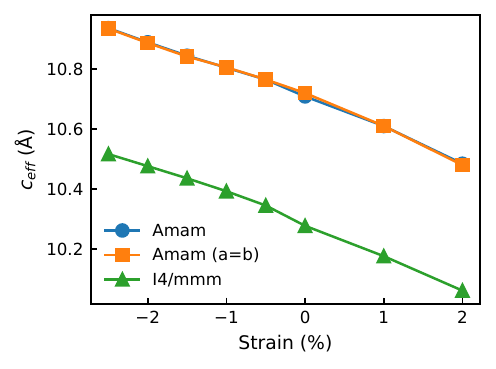}
\includegraphics[width=0.8\columnwidth]{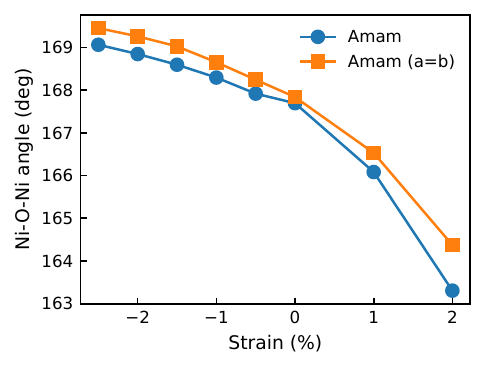}
\caption{Evolution of the out-of-plane lattice parameter $c$ (top) and the in-plane Ni--O--Ni bond angle (bottom) for the strained structures. The out-of-plane parameter decreases as strain is tuned from compressive to tensile values, while the bond angle decreases with increasing tensile strain. The bond angle for the $I4/mmm$ phase (not shown) is fixed at 180$^\circ$ by symmetry.}
\label{fig:structure}
\end{figure}

Figure~\ref{fig:structure} shows the evolution of the out-of-plane lattice parameter $c$ and in-plane Ni--O--Ni bond angle. $c$ decreases linearly from compressive to tensile strain, consistent with a Poisson response. As expected, the Ni--O--Ni bond angle decreases with tensile strain, indicating enhanced octahedral tilting, while compressive strain straightens the bonds~\cite{bhattaStructuralElectronicEvolution2025, bhattResolvingStructuralOrigins2025}. $Amam$ and $Amam$ ($a = b$) trends are similar, with only modest differences throughout. In $I4/mmm$, the angle is fixed at 180$^\circ$ by symmetry. Consistent with Ref.~\cite{bhattaStructuralElectronicEvolution2025}, the $Amam$ bond angles do not fully straighten to 180$^\circ$ even under large compression.

\subsection{Electronic structure and low-energy spectral weight}

\subsubsection{Wannier and DFT band structures}
\begin{figure}[t]
\centering
\includegraphics[width=0.3\columnwidth]{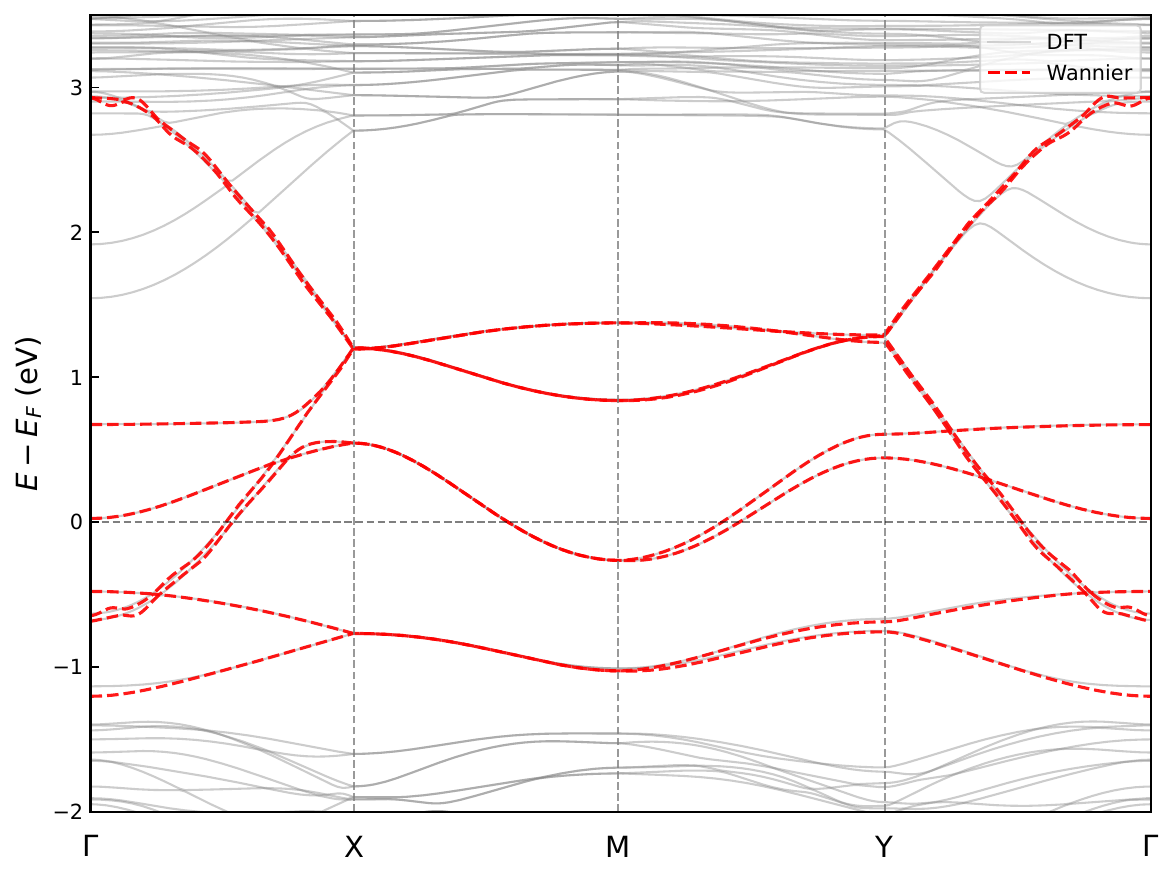}
\includegraphics[width=0.3\columnwidth]{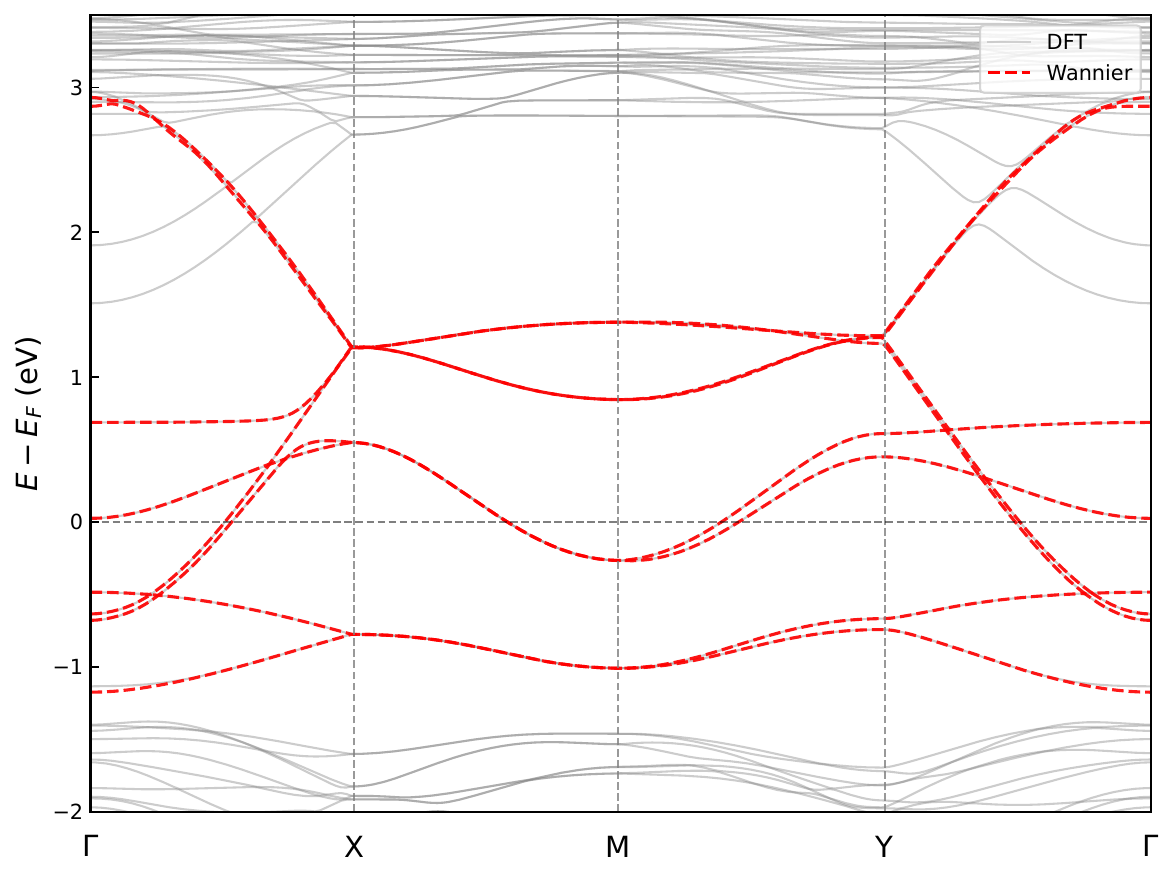}
\includegraphics[width=0.3\columnwidth]{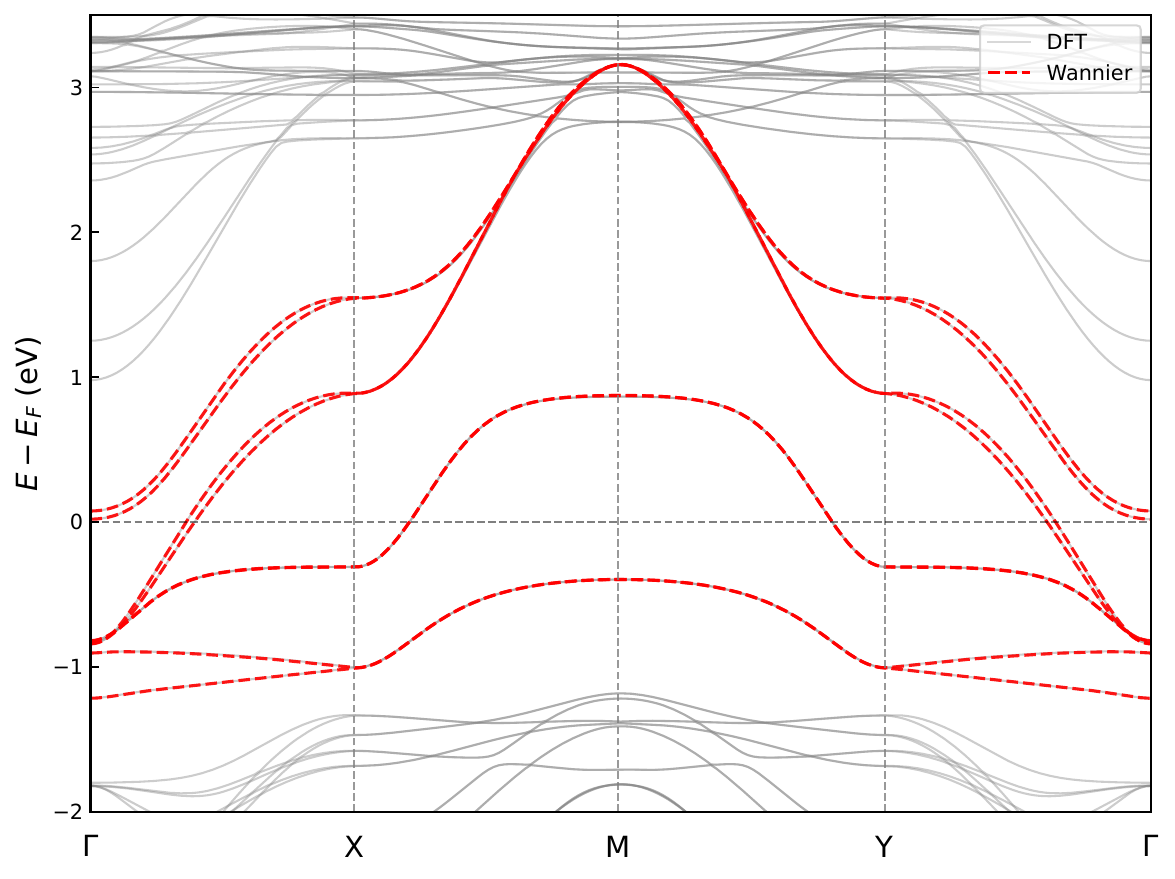}
\caption{Representative validation at $-2.0\%$ strain, comparing DFT bands (black) to the Wannier interpolation (red dashed) near the Fermi level for $Amam$ (left), $Amam$ ($a = b$) (center), and $I4/mmm$ (right).}
\label{fig:bands}
\end{figure}

Figure~\ref{fig:bands} compares DFT and Wannier bands at $-2.0\%$ strain, showing excellent agreement near the Fermi level. A gap around $-1.2$ eV separates bonding states from the Ni $e_g$ manifold. 

\begin{figure*}[t]
\centering
\includegraphics[width=0.6\columnwidth]{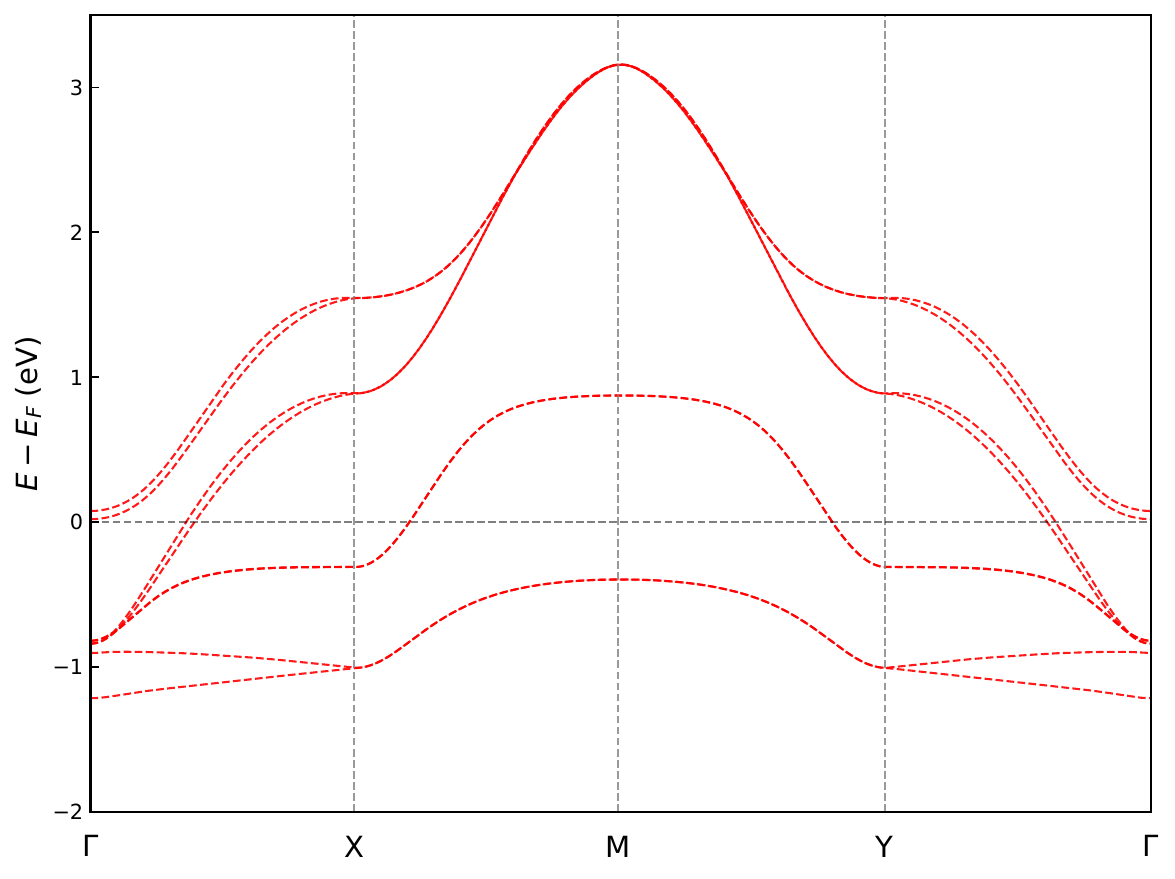}
\includegraphics[width=0.6\columnwidth]{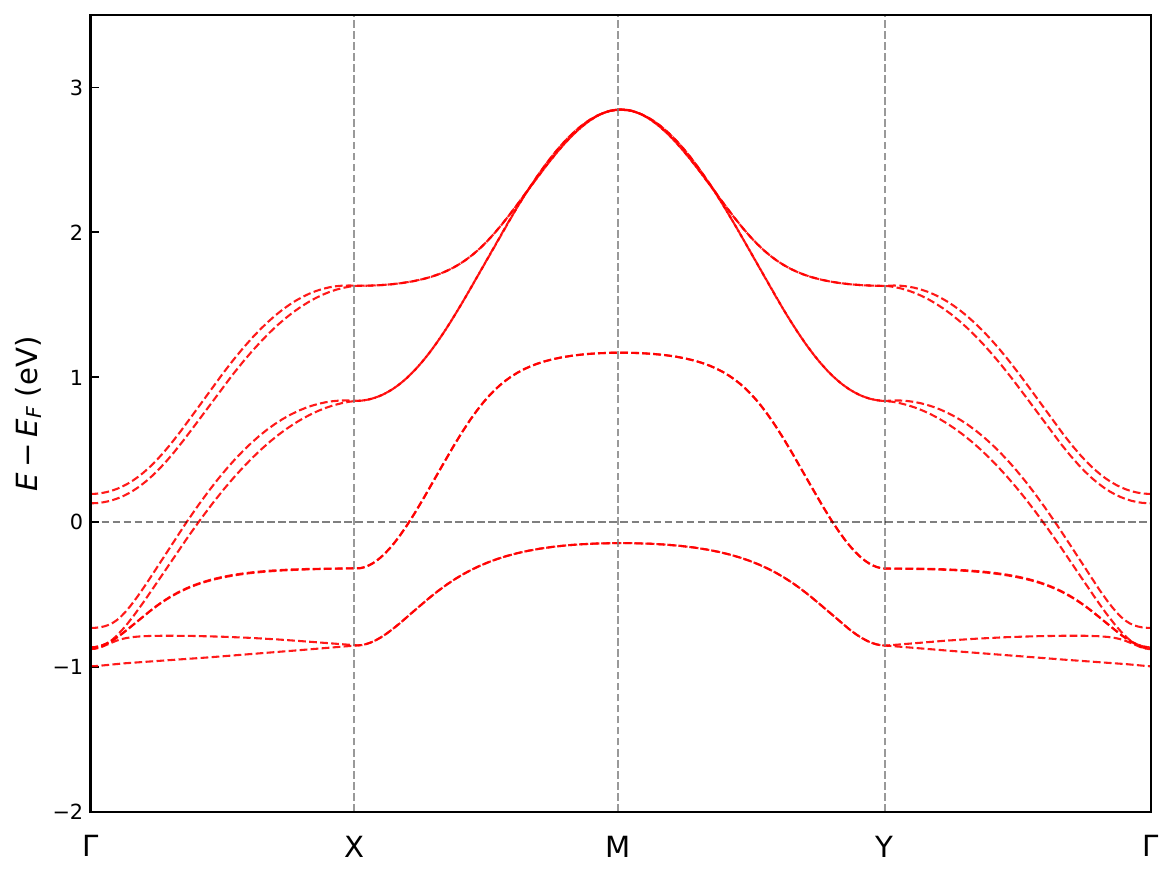}
\includegraphics[width=0.6\columnwidth]{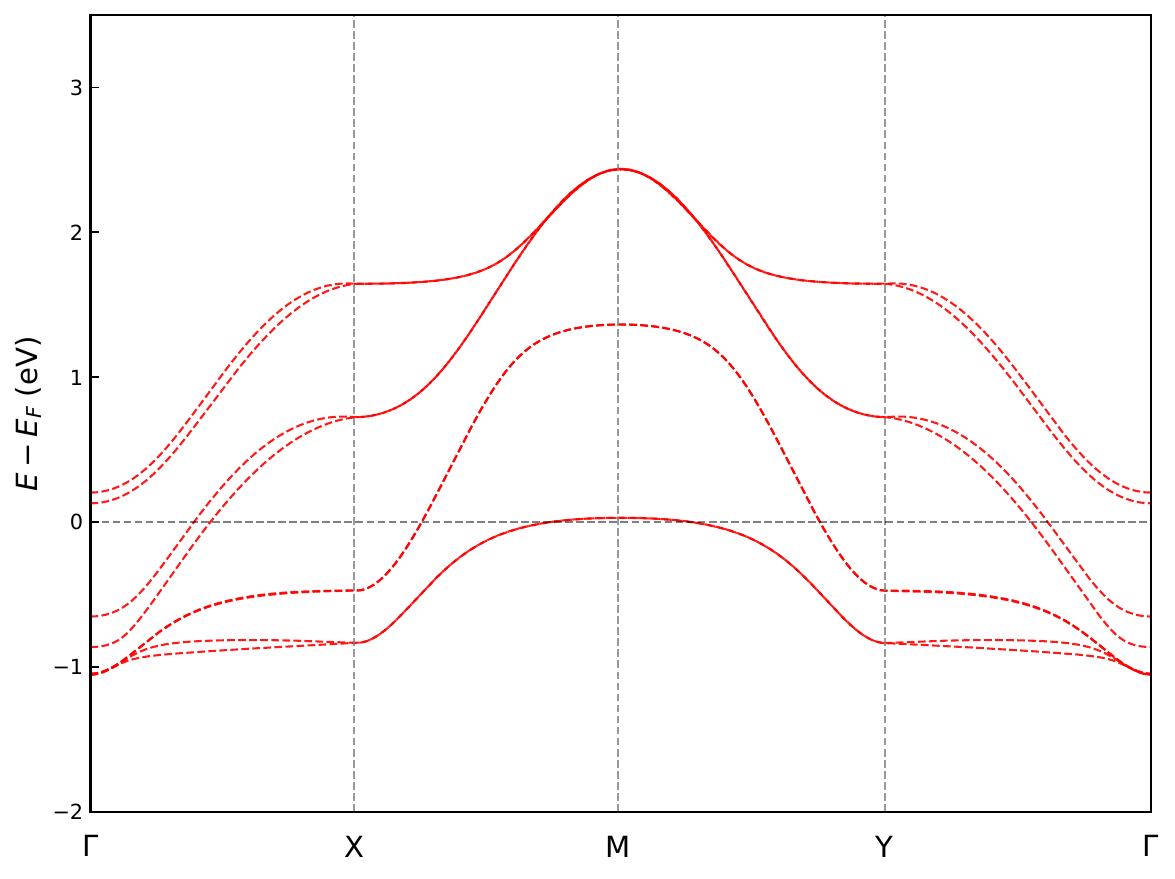} \\
\includegraphics[width=0.6\columnwidth]{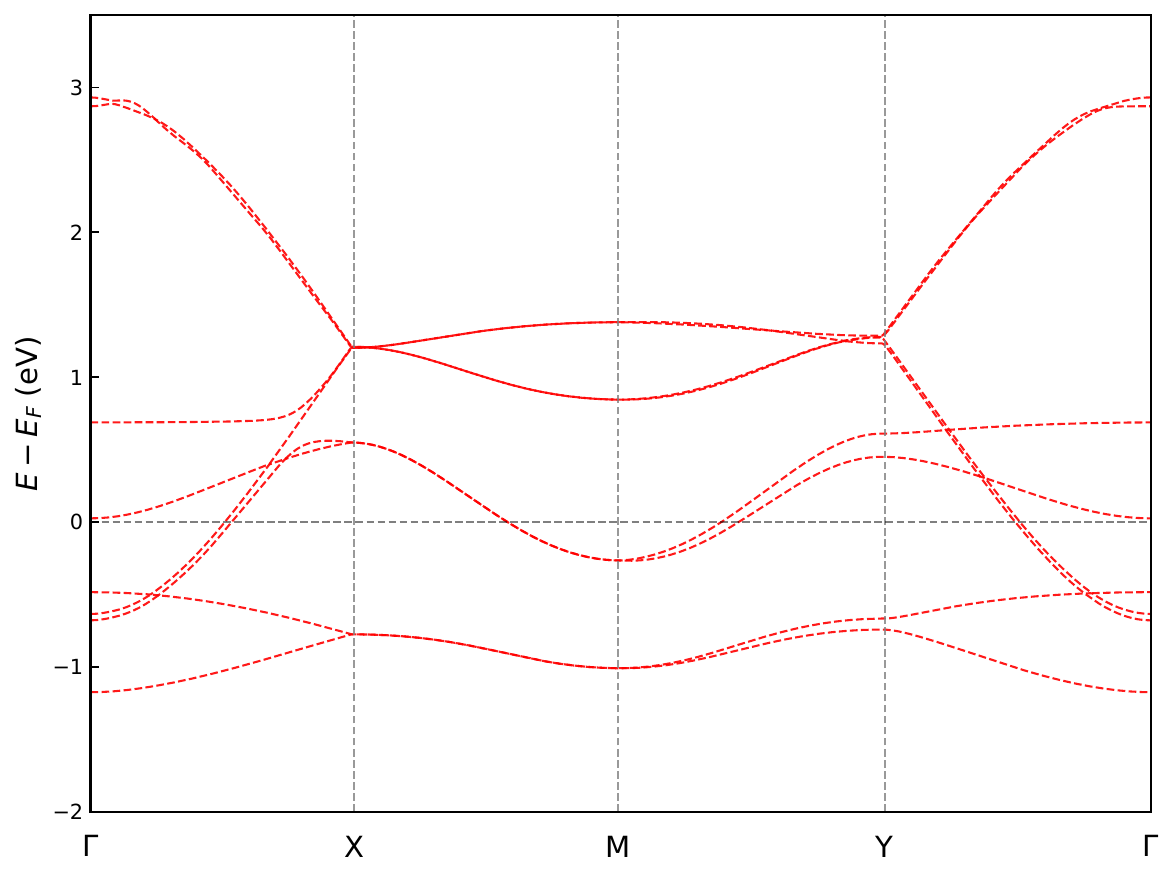}
\includegraphics[width=0.6\columnwidth]{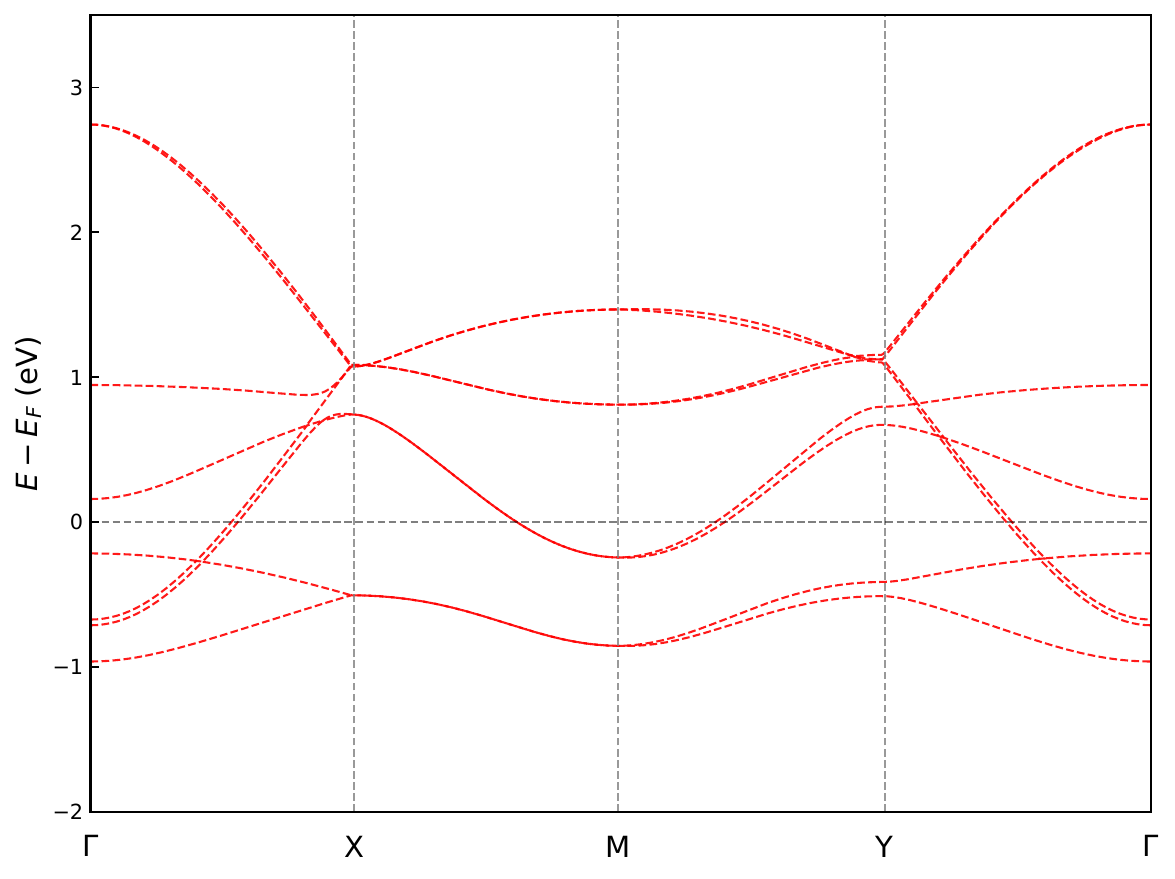}
\includegraphics[width=0.6\columnwidth]{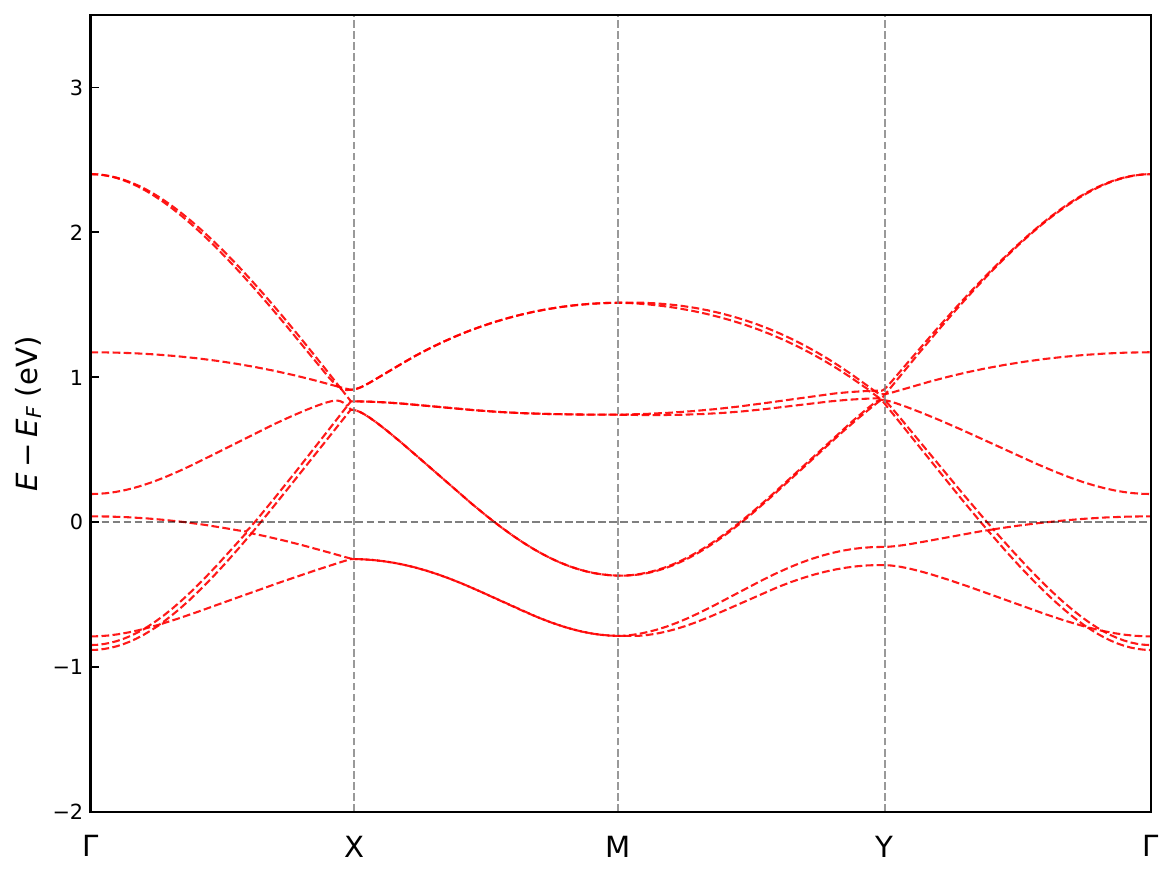}
\caption{Wannier band structures for the $I4/mmm$ phase (top row) and $Amam$ ($a=b$) phase (bottom row) under biaxial strain. Columns correspond to $-2.0\%$ (compressive), $0\%$ (unstrained), and $+2.0\%$ (tensile) strain. The Fermi level is at 0 eV. Note that because of the skewed and rotated $Amam$ cell, the bands do not cleanly align with the $I4/mmm$ bands.}
\label{fig:wan_bands_evolution}
\end{figure*}

Figure~\ref{fig:wan_bands_evolution} shows the Wannier band evolution. Note that $Amam$ calculations used a rotated supercell, folding the $I4/mmm$ zone-corner M point to $\Gamma$. The $d_{x^2-y^2}$ bands are identified by their large in-plane bandwidth, while the $d_{z^2}$ bands appear as bonding--antibonding pairs with comparable in-plane dispersion but sizable energy splitting arising from bilayer coupling.

\subsubsection{Density of states at the Fermi level}
\begin{figure}[t]
\centering
\includegraphics[width=0.95\columnwidth]{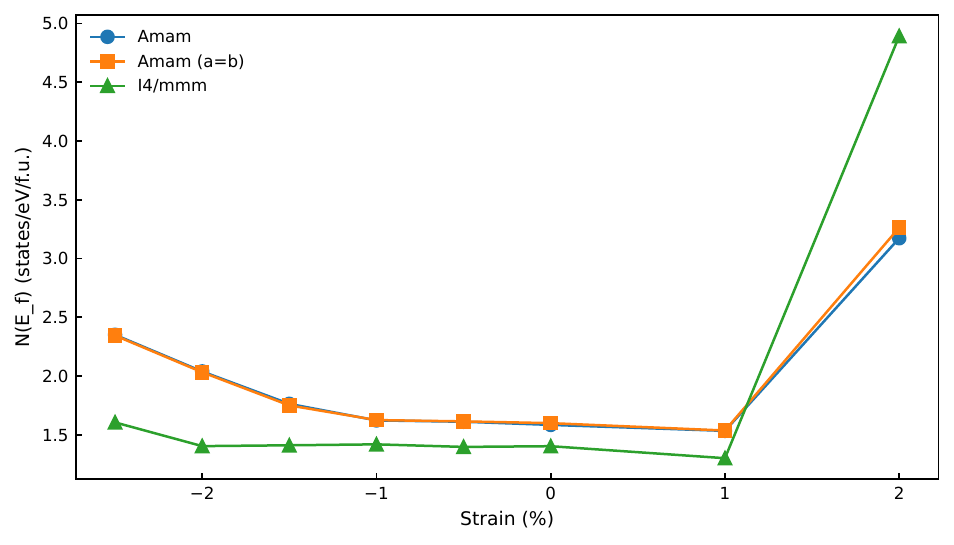}
\caption{Strain dependence of the density of states at the Fermi level, $N(E_F)$. A nonmonotonic evolution is observed, with a sharp rise at large tensile strain.}
\label{fig:dos}
\end{figure}

Figure~\ref{fig:dos} shows the strain dependence of the density of states at the Fermi level, $N(E_F)$. $N(E_F)$ evolves nonmonotonically, with a sharp rise at large tensile strain across all phases. This increase is likely driven by a Lifshitz transition where a Ni $d_{z^2}$ bonding band crosses the Fermi level (at M in $I4/mmm$, $\Gamma$ in $Amam$), as seen in Figure~\ref{fig:wan_bands_evolution}. There is also a modest increase in $N(E_F)$ under compressive strain for the $Amam$ phases, likely due to a different band approaching $E_f$ near $\Gamma$.

\subsubsection{Fermi surface evolution}
\begin{figure}[t]
\centering
\includegraphics[width=\columnwidth]{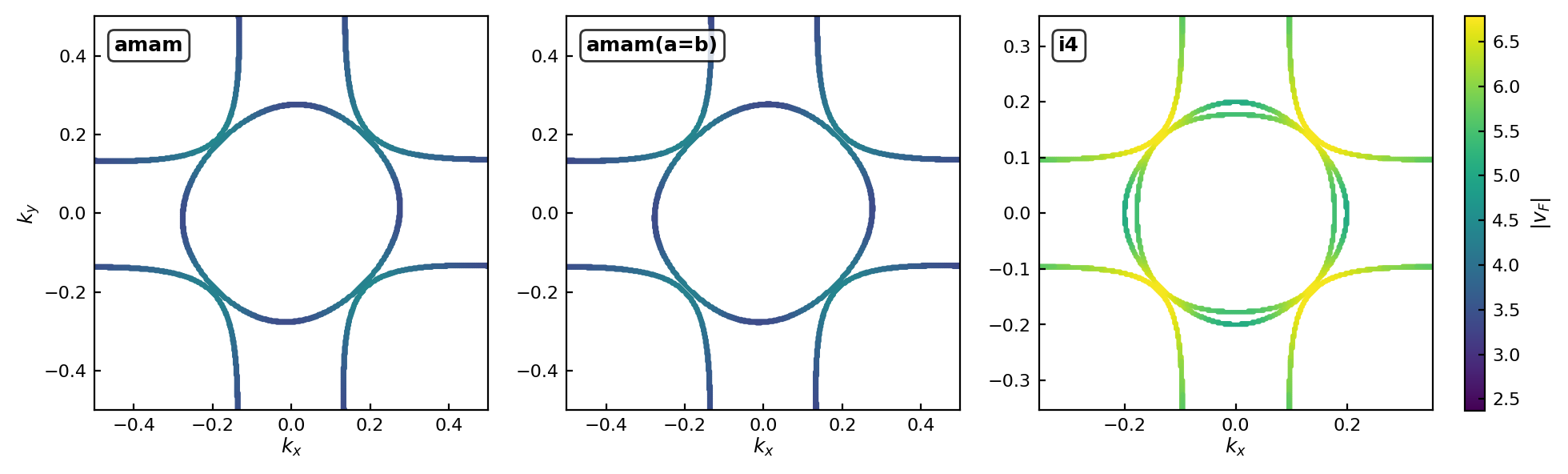}
\caption{Fermi surface comparison at 0\% strain for $Amam$ (left), $Amam$ ($a=b$) (center), and $I4/mmm$ (right). Note that the $I4/mmm$ phase has been zoomed in to match the $Amam$ supercell, and that the $Amam$ phases have been rotated to align with the $I4/mmm$ axes.}
\label{fig:fs_comp}
\end{figure}

\begin{figure}[t]
\centering
\includegraphics[width=\columnwidth]{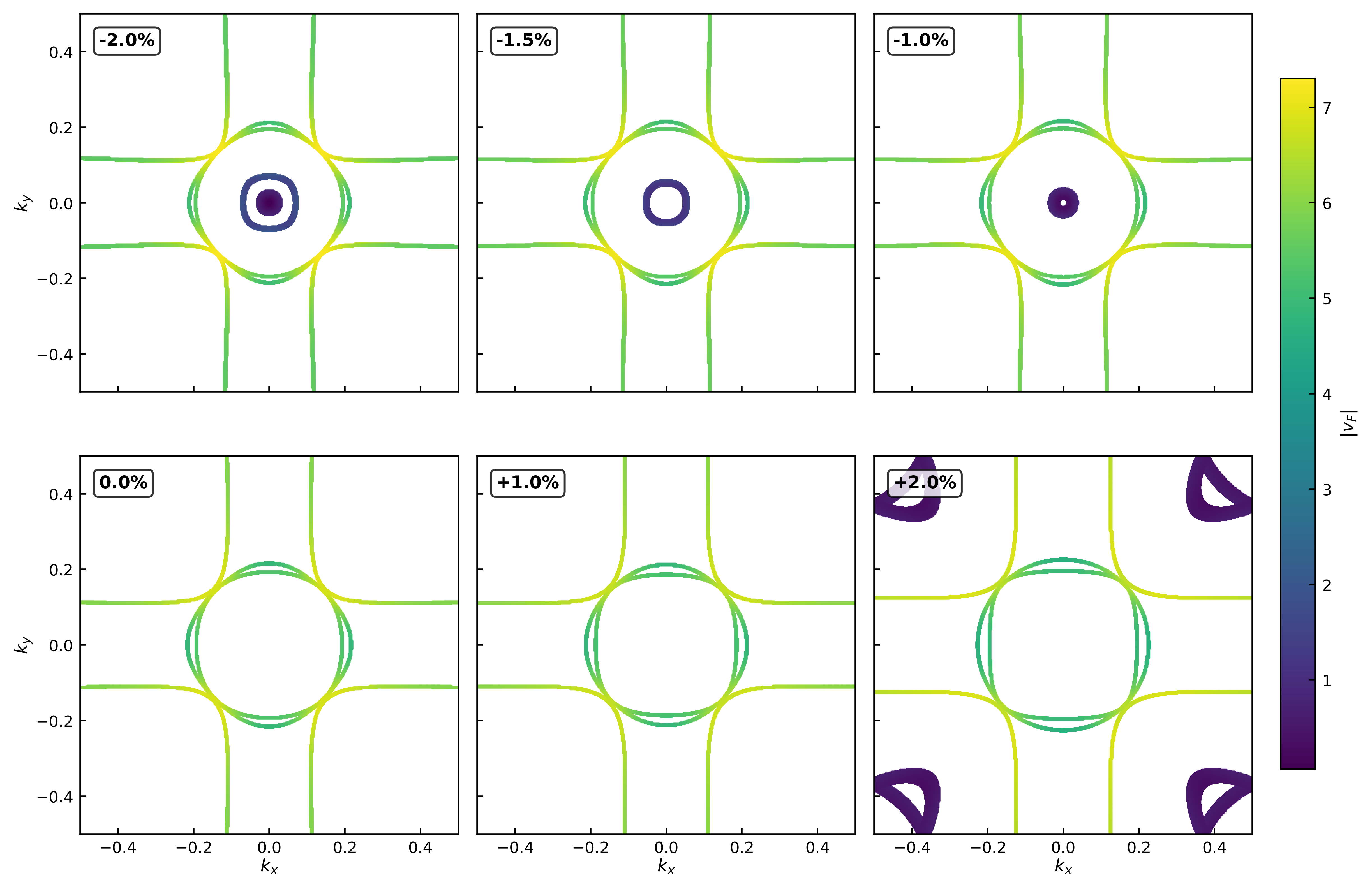}
\caption{Fermi surface evolution of the $I4/mmm$ phase under selected strains. Under compressive strain, a small $\Gamma$-centered pocket appears due to a strain-induced band crossing at the Fermi level. This pocket vanishes near zero strain. At $+2.0\%$ tensile strain, a Lifshitz transition occurs, with new pockets emerging near the Brillouin-zone corners (M points).}
\label{fig:fs_i4}
\end{figure}

At zero strain, all phases show a central pocket and quasi-1D sheets (Fig.~\ref{fig:fs_comp}). Under compression, $I4/mmm$ develops a small $\Gamma$-centered pocket (Fig.~\ref{fig:fs_i4}), arising from a shift of the chemical potential at fixed filling that drives a $d_{z^2}$-derived band to cross the Fermi level. Under large tensile strain, all phases undergo a Lifshitz transition, with new pockets emerging near the zone corners (M points).

\subsection{Magnetic proximity under biaxial strain}
\begin{figure}[t]
\centering
\includegraphics[width=0.95\columnwidth]{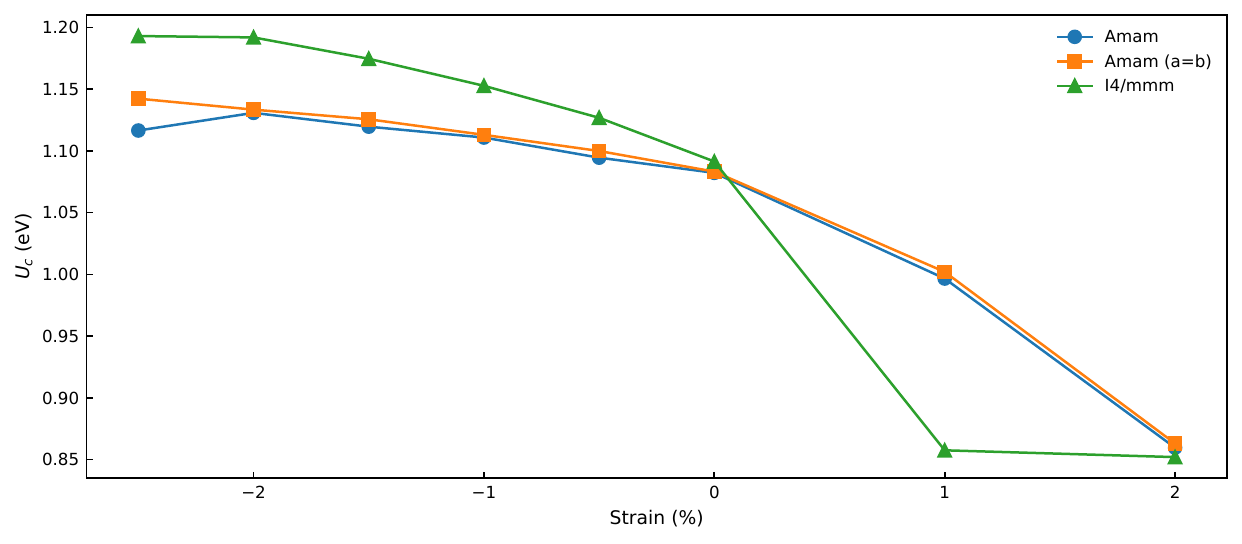}
\caption{Strain dependence of $U_c$, defined as the interaction strength $U$ at which the leading magnetic eigenmode reaches the Stoner instability condition. Smaller $U_c$ corresponds to stronger magnetic tendency.}
\label{fig:mag}
\end{figure}

Figure~\ref{fig:mag} shows the critical interaction strength $U_c$ for magnetic instability (defined so that the largest eigenvalue of $U_s \chi_0(\mathbf{q})$ at $U = U_c$ equals 1). $U_c$ generally decreases from compressive to tensile strain, indicating enhanced magnetic proximity under tensile strains, which are likely closer to spin-density-wave instabilities~\cite{sunSignaturesSuperconductivity802023, liuEvidenceChargeSpin2022}.

\subsection{$T_c$ evolution under biaxial strain}
\begin{figure}[t]
\centering
\includegraphics[width=0.8\columnwidth]{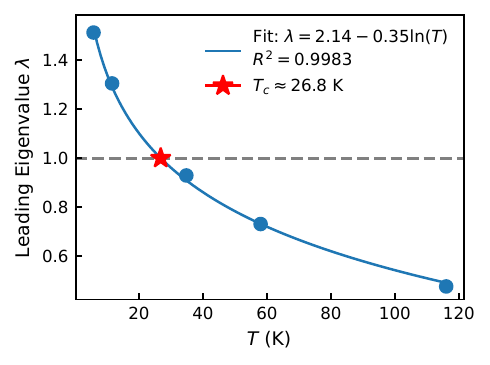}
\caption{Temperature dependence of the leading pairing eigenvalue $\lambda$ for the $I4/mmm$ phase at $-2.0\%$ strain. The solid line represents a logarithmic fit $\lambda = A + B \ln(T)$, used to determine $T_c$ at $\lambda=1$.}
\label{fig:lambda_vs_T}
\end{figure}

\begin{figure}[t]
\centering
\includegraphics[width=\columnwidth]{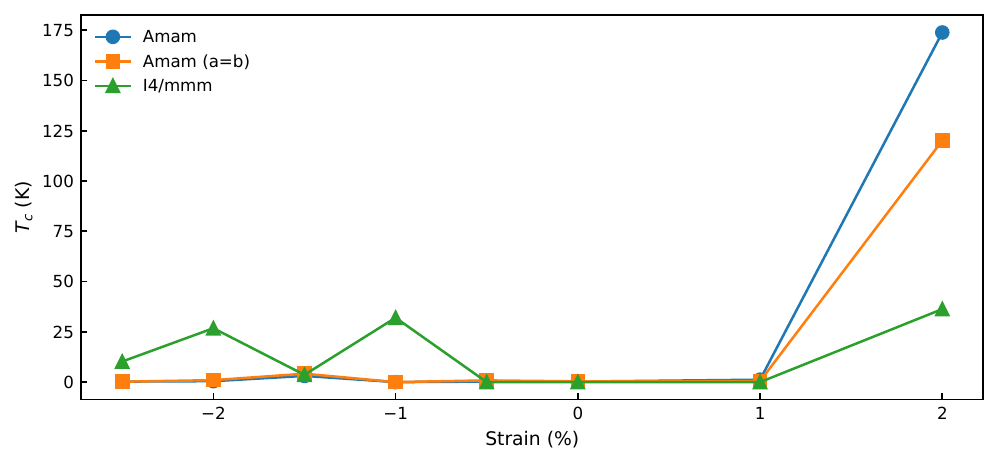}
\caption{$T_c$ vs. strain for $Amam$, $Amam$ ($a=b$), and $I4/mmm$ phases. A pronounced enhancement of $T_c$ is observed at large tensile strains for all structural paths, coinciding with a Lifshitz transition in the Fermi surface. In addition, the $I4/mmm$ phase exhibits a moderate $T_c$ enhancement under certain compressive strains.}
\label{fig:Tc_vs_strain}
\end{figure}

Figure~\ref{fig:lambda_vs_T} illustrates $T_c$ determination for $I4/mmm$ at $-2.0\%$ strain. Figure~\ref{fig:Tc_vs_strain} summarizes $T_c$ vs. strain. Large tensile strain strongly enhances $T_c$ across all phases, coinciding with the Lifshitz transition and increased spectral weight. This large value is likely an overestimate caused by limitations of the model (e.g. neglected strong-coupling effects and competing order), but qualitatively agrees with prior theoretical works~\cite{lechermannElectronicCorrelationsSuperconducting2023, luInterlayerCouplingDrivenHighTemperatureSuperconductivity2024, geislerFermiSurfaceReconstruction2025}. $I4/mmm$ also shows moderate enhancement under compression ($\approx$32 K at -1.0\%, $\approx$27 K at -2.0\%), comparable to experimental values ($\approx$26-42 K)~\cite{koSignaturesAmbientPressure2025}. This suggests our framework captures key aspects of strain-dependent superconductivity in \lno.

\subsubsection{Robustness of results}
We tested robustness against variations in Hubbard $U$, k-mesh density, smearing, Wannier windows, and filling. The trends of tensile enhancement and, for $I4/mmm$, compressive increase, remained consistent, suggesting that they are robust features of the model.

\section{Discussion}
Our results demonstrate that strain acts as a tuning parameter that changes the balance between structural symmetry, electronic topology, and magnetic proximity, rather than simply modifying bandwidths. The small energy separation between the $Amam$, $Amam$ ($a=b$), and $I4/mmm$ phases (Fig.~\ref{fig:energy}) indicates that multiple structural phases are experimentally accessible, particularly in thin-film geometries, and that superconductivity does not strictly require the global energy minimum structure, consistent with observations of mixed-phase samples~\cite{puphalUnconventionalCrystalStructure2024, wangStructureResponsibleSuperconducting2024, zhangStructuralPhaseTransition2024}. Notably, the near-degeneracy of the $Amam$ and $Amam$ ($a=b$) phases suggests that enforcing in-plane tetragonality alone has little effect on the pairing tendency, implying that static crystal symmetry is not a dominant factor. At the same time, compressive strain significantly lowers the energetic penalty of the high-symmetry $I4/mmm$ phase while straightening the Ni--O--Ni bonds (Fig.~\ref{fig:structure}), thereby enhancing interlayer superexchange and stabilizing electronic features associated with pressure-induced superconductivity, as discussed in Refs.~\cite{bhattaStructuralElectronicEvolution2025, bhattResolvingStructuralOrigins2025}.

Electronically, strain primarily tunes the relative position and occupation of the Ni $d_{z^2}$ states, while the $d_{x^2-y^2}$ bands remain broadly dispersive and quasi-two-dimensional across all strain values (Fig.~\ref{fig:wan_bands_evolution}). Under tensile strain, a $d_{z^2}$-derived bonding band crosses the Fermi level, producing a Lifshitz transition that sharply increases the low-energy spectral weight, similar to findings in Refs.~\cite{geislerFermiSurfaceReconstruction2025, shaoBandStructurePairing2025}. Consequently, the density of states at the Fermi level, $N(E_F)$, exhibits a nonmonotonic evolution directly tied to changes in Fermi surface topology rather than gradual band renormalization (Fig.~\ref{fig:dos}). While the specific location of the emergent pocket depends on crystal symmetry (M point in $I4/mmm$, $\Gamma$ in folded $Amam$), its physical role of introducing additional $d_{z^2}$ character near the Fermi level is consistent across phases. Conversely, under compressive strain in the $I4/mmm$ phase, a small $\Gamma$-centered pocket appears (Fig.~\ref{fig:fs_i4}), leading to a more modest enhancement of $N(E_F)$. Notably, this pocket arises from a shift in chemical potential at fixed filling, suggesting that doping may be an interesting pathway to further tune superconductivity.

Tensile strain strongly enhances $T_c$ in our calculations by driving a Lifshitz transition that increases low-energy spectral weight and amplifies spin fluctuations through improved nesting conditions, consistent with theoretical proposals for the bulk material~\cite{lechermannElectronicCorrelationsSuperconducting2023, luInterlayerCouplingDrivenHighTemperatureSuperconductivity2024}. However, tensile strain also produces a large reduction in the critical interaction strength $U_c$ (Fig.~\ref{fig:mag}), indicating increased magnetic proximity and likely bringing the system close to a competing spin-density-wave instability, in agreement with reports of magnetic order in the parent compounds~\cite{chenElectronicMagneticExcitations2024, liuEvidenceChargeSpin2022}. As a result, superconductivity under tensile strain may be strongly constrained by competition with magnetic order, potentially explaining why tensile-strained samples do not exhibit superconductivity in experiment~\cite{osadaStraintuningSuperconductivityLa3Ni2O72025} despite the large pairing strength predicted here and in other theoretical studies~\cite{geislerFermiSurfaceReconstruction2025}. The similar $T_c$ trends for $Amam$ and $Amam$ ($a=b$) further reinforce that superconductivity is primarily governed by electronic structure evolution rather than structural details. 

Crucially, the moderate $T_c$ enhancement found under compressive strain in the $I4/mmm$ phase aligns with experimental observations in strained thin films~\cite{koSignaturesAmbientPressure2025, zhouAmbientpressureSuperconductivityOnset2025, osadaStraintuningSuperconductivityLa3Ni2O72025}. This effect is structurally selective: in the high-symmetry $I4/mmm$ phase, straight Ni--O--Ni bonds allow compression to enhance interlayer coupling, driving a $d_{z^2}$-derived band across the chemical potential. This creates a small $\Gamma$-centered hole pocket (Fig.~\ref{fig:fs_i4}), appearing in our calculations due to the fixed filling constraint ($n=1.5$)—which increases low-energy spectral weight and spin fluctuations. Conversely, octahedral tilting in the $Amam$ phases limits interlayer coupling, preventing the formation of this pocket and suppressing $T_c$ enhancement. These findings support a spin-fluctuation-mediated pairing mechanism where $d_{z^2}$ metallization, while not necessarily required, significantly boosts pairing, consistent with recent work on the $\gamma$ pocket~\cite{geislerFermiSurfaceReconstruction2025, shaoBandStructurePairing2025}. Overall, the agreement between our calculated and experimental $T_c$ values suggests that our framework captures the essential physics of strain-tuned superconductivity in \lno.
\linebreak
\section{Conclusion}
In summary, we have theoretically investigated the strain dependence of superconductivity in \lno\ using a combined DFT+$U$ and RPA spin-fluctuation framework. Our results reveal a nonmonotonic evolution of $T_c$ driven by the interactions between structural symmetry, electronic topology, and magnetic correlations. We find that large tensile strain induces a Lifshitz transition, which is characterized by a Ni $d_{z^2}$ band crossing the Fermi level. This transition sharply increases the density of states and spin fluctuations, theoretically enhancing $T_c$. However, this regime is also accompanied by a strong increase in magnetic proximity, suggesting that competing spin-density-wave order may suppress superconductivity in experiments.

Under compressive strain, we identify a structurally selective enhancement of $T_c$ specific to the high-symmetry $I4/mmm$ phase. This effect arises from the formation of a small $\Gamma$-centered hole pocket ($\gamma$ pocket), boosting interlayer coupling. The resulting moderate $T_c$ enhancement aligns well with experimental observations in compressively strained thin films, highlighting the critical role of the $\gamma$ pocket in optimizing pairing. Our findings demonstrate that strain is an effective tuning parameter for nickelate superconductivity, capable of manipulating Fermi surface topology and magnetic competition to access higher transition temperatures. These insights suggest that stabilizing the $I4/mmm$ phase under moderate compressive strain or doping to tune the $\gamma$ pocket occupancy offer promising routes for optimizing ambient-pressure superconductivity in \lno.

\section{Data Availability}
The data and code supporting the findings of this study are available from the author upon reasonable request.

\bibliographystyle{apsrev4-2}
\bibliography{refs}

@article{anisimovBand1991,
  title = {Band Theory and {{Mott}} Insulators: {{Hubbard U}} Instead of {{Stoner I}}},
  shorttitle = {Band Theory and {{Mott}} Insulators},
  author = {Anisimov, Vladimir I. and Zaanen, Jan and Andersen, Ole K.},
  year = 1991,
  month = jul,
  journal = {Physical Review B},
  volume = {44},
  number = {3},
  pages = {943--954},
  publisher = {American Physical Society},
  doi = {10.1103/PhysRevB.44.943},
  urldate = {2025-12-14}
}

@article{anisimovElectronicStructurePossible1999,
  title = {Electronic Structure of Possible Nickelate Analogs to the Cuprates},
  author = {Anisimov, V. I. and Bukhvalov, D. and Rice, T. M.},
  year = 1999,
  month = mar,
  journal = {Physical Review B},
  volume = {59},
  number = {12},
  pages = {7901--7906},
  publisher = {American Physical Society},
  doi = {10.1103/PhysRevB.59.7901},
  urldate = {2025-12-16}
}

@misc{bhattaStructuralElectronicEvolution2025,
  title = {Structural and {{Electronic Evolution}} of {{Bilayer Nickelates Under Biaxial Strain}}},
  author = {Bhatta, H. C. Regan B. and Zhang, Xiaoliang and Zhong, Yong and Jia, Chunjing},
  year = 2025,
  month = feb,
  number = {arXiv:2502.01624},
  eprint = {2502.01624},
  primaryclass = {cond-mat},
  publisher = {arXiv},
  doi = {10.48550/arXiv.2502.01624},
  urldate = {2025-11-29},
  archiveprefix = {arXiv}
}

@misc{bhattResolvingStructuralOrigins2025,
  title = {Resolving {{Structural Origins}} for {{Superconductivity}} in {{Strain-Engineered La}}\$\_3\${{Ni}}\$\_2\${{O}}\$\_7\$ {{Thin Films}}},
  author = {Bhatt, Lopa and Jiang, Abigail Y. and Ko, Eun Kyo and Schnitzer, Noah and Pan, Grace A. and Segedin, Dan Ferenc and Liu, Yidi and Yu, Yijun and Zhao, Yi-Feng and Morales, Edgar Abarca and Brooks, Charles M. and Botana, Antia S. and Hwang, Harold Y. and Mundy, Julia A. and Muller, David A. and Goodge, Berit H.},
  year = 2025,
  month = jan,
  number = {arXiv:2501.08204},
  eprint = {2501.08204},
  primaryclass = {cond-mat},
  publisher = {arXiv},
  doi = {10.48550/arXiv.2501.08204},
  urldate = {2025-12-07},
  archiveprefix = {arXiv}
}

@article{chenElectronicMagneticExcitations2024,
  title = {Electronic and Magnetic Excitations in {{La3Ni2O7}}},
  author = {Chen, Xiaoyang and Choi, Jaewon and Jiang, Zhicheng and Mei, Jiong and Jiang, Kun and Li, Jie and Agrestini, Stefano and {Garcia-Fernandez}, Mirian and Sun, Hualei and Huang, Xing and Shen, Dawei and Wang, Meng and Hu, Jiangping and Lu, Yi and Zhou, Ke-Jin and Feng, Donglai},
  year = 2024,
  month = nov,
  journal = {Nature Communications},
  volume = {15},
  number = {1},
  pages = {9597},
  issn = {2041-1723},
  doi = {10.1038/s41467-024-53863-5},
  urldate = {2025-11-06},
  langid = {english}
}

@article{chenEvidenceSpinDensity2024,
  title = {Evidence of {{Spin Density Waves}} in \$\textbraceleft\textbackslash mathrm\textbraceleft{{La}}\textbraceright\textbraceright\_\textbraceleft 3\textbraceright\textbraceleft\textbackslash mathrm\textbraceleft{{Ni}}\textbraceright\textbraceright\_\textbraceleft 2\textbraceright\textbraceleft\textbackslash mathrm\textbraceleft{{O}}\textbraceright\textbraceright\_\textbraceleft 7\textbackslash ensuremath\textbraceleft -\textbraceright\textbackslash ensuremath\textbraceleft\textbackslash delta\textbraceright\textbraceright\$},
  author = {Chen, Kaiwen and Liu, Xiangqi and Jiao, Jiachen and Zou, Muyuan and Jiang, Chengyu and Li, Xin and Luo, Yixuan and Wu, Qiong and Zhang, Ningyuan and Guo, Yanfeng and Shu, Lei},
  year = 2024,
  month = jun,
  journal = {Physical Review Letters},
  volume = {132},
  number = {25},
  pages = {256503},
  publisher = {American Physical Society},
  doi = {10.1103/PhysRevLett.132.256503},
  urldate = {2025-12-15}
}

@article{christianssonCorrelatedElectronicStructure2023,
  title = {Correlated {{Electronic Structure}} of {{La}} 3 {{Ni}} 2 {{O}} 7 under {{Pressure}}},
  author = {Christiansson, Viktor and Petocchi, Francesco and Werner, Philipp},
  year = 2023,
  month = nov,
  journal = {Physical Review Letters},
  volume = {131},
  number = {20},
  pages = {206501},
  issn = {0031-9007, 1079-7114},
  doi = {10.1103/PhysRevLett.131.206501},
  urldate = {2025-11-21},
  langid = {english}
}

@article{dudarevElectronenergylossSpectraStructural1998,
  title = {Electron-Energy-Loss Spectra and the Structural Stability of Nickel Oxide: {{An LSDA}}+{{U}} Study},
  shorttitle = {Electron-Energy-Loss Spectra and the Structural Stability of Nickel Oxide},
  author = {Dudarev, S. L. and Botton, G. A. and Savrasov, S. Y. and Humphreys, C. J. and Sutton, A. P.},
  year = 1998,
  month = jan,
  journal = {Physical Review B},
  volume = {57},
  number = {3},
  pages = {1505--1509},
  issn = {0163-1829, 1095-3795},
  doi = {10.1103/PhysRevB.57.1505},
  urldate = {2025-11-06},
  copyright = {http://link.aps.org/licenses/aps-default-license},
  langid = {english}
}

@article{garrityPseudopotentialsHighthroughputDFT2014,
  title = {Pseudopotentials for High-Throughput {{DFT}} Calculations},
  author = {Garrity, Kevin F. and Bennett, Joseph W. and Rabe, Karin M. and Vanderbilt, David},
  year = 2014,
  month = jan,
  journal = {Computational Materials Science},
  volume = {81},
  pages = {446--452},
  issn = {09270256},
  doi = {10.1016/j.commatsci.2013.08.053},
  urldate = {2025-11-03},
  langid = {english}
}

@misc{geislerFermiSurfaceReconstruction2025,
  title = {Fermi Surface Reconstruction and Enhanced Spin Fluctuations in Strained {{La}}\$\_3\${{Ni}}\$\_2\${{O}}\$\_\textbraceleft 7\textbraceright\$ on {{LaAlO}}\$\_3\$(001) and {{SrTiO}}\$\_3\$(001)},
  author = {Geisler, Benjamin and Hamlin, James J. and Stewart, Gregory R. and Hennig, Richard G. and Hirschfeld, P. J.},
  year = 2025,
  month = feb,
  number = {arXiv:2411.14600},
  eprint = {2411.14600},
  primaryclass = {cond-mat},
  publisher = {arXiv},
  doi = {10.48550/arXiv.2411.14600},
  urldate = {2025-11-29},
  archiveprefix = {arXiv}
}

@article{giannozziAdvancedCapabilitiesMaterials2017,
  title = {Advanced Capabilities for Materials Modelling with {{Quantum ESPRESSO}}},
  author = {Giannozzi, P and Andreussi, O and Brumme, T and Bunau, O and Buongiorno Nardelli, M and Calandra, M and Car, R and Cavazzoni, C and Ceresoli, D and Cococcioni, M and Colonna, N and Carnimeo, I and Dal Corso, A and De Gironcoli, S and Delugas, P and DiStasio, R A and Ferretti, A and Floris, A and Fratesi, G and Fugallo, G and Gebauer, R and Gerstmann, U and Giustino, F and Gorni, T and Jia, J and Kawamura, M and Ko, H-Y and Kokalj, A and K{\"u}{\c c}{\"u}kbenli, E and Lazzeri, M and Marsili, M and Marzari, N and Mauri, F and Nguyen, N L and Nguyen, H-V and {Otero-de-la-Roza}, A and Paulatto, L and Ponc{\'e}, S and Rocca, D and Sabatini, R and Santra, B and Schlipf, M and Seitsonen, A P and Smogunov, A and Timrov, I and Thonhauser, T and Umari, P and Vast, N and Wu, X and Baroni, S},
  year = 2017,
  month = nov,
  journal = {Journal of Physics: Condensed Matter},
  volume = {29},
  number = {46},
  pages = {465901},
  issn = {0953-8984, 1361-648X},
  doi = {10.1088/1361-648X/aa8f79},
  urldate = {2025-11-03}
}

@article{giannozziQUANTUMESPRESSOModular2009,
  title = {{{QUANTUM ESPRESSO}}: A Modular and Open-Source Software Project for Quantum Simulations of Materials},
  shorttitle = {{{QUANTUM ESPRESSO}}},
  author = {Giannozzi, Paolo and Baroni, Stefano and Bonini, Nicola and Calandra, Matteo and Car, Roberto and Cavazzoni, Carlo and Ceresoli, Davide and Chiarotti, Guido L and Cococcioni, Matteo and Dabo, Ismaila and Dal Corso, Andrea and De Gironcoli, Stefano and Fabris, Stefano and Fratesi, Guido and Gebauer, Ralph and Gerstmann, Uwe and Gougoussis, Christos and Kokalj, Anton and Lazzeri, Michele and {Martin-Samos}, Layla and Marzari, Nicola and Mauri, Francesco and Mazzarello, Riccardo and Paolini, Stefano and Pasquarello, Alfredo and Paulatto, Lorenzo and Sbraccia, Carlo and Scandolo, Sandro and Sclauzero, Gabriele and Seitsonen, Ari P and Smogunov, Alexander and Umari, Paolo and Wentzcovitch, Renata M},
  year = 2009,
  month = sep,
  journal = {Journal of Physics: Condensed Matter},
  volume = {21},
  number = {39},
  pages = {395502},
  issn = {0953-8984, 1361-648X},
  doi = {10.1088/0953-8984/21/39/395502},
  urldate = {2025-11-03}
}

@article{graserNeardegeneracySeveralPairing2009,
  title = {Near-Degeneracy of Several Pairing Channels in Multiorbital Models for the {{Fe}} Pnictides},
  author = {Graser, S and Maier, T A and Hirschfeld, P J and Scalapino, D J},
  year = 2009,
  month = feb,
  journal = {New Journal of Physics},
  volume = {11},
  number = {2},
  pages = {025016},
  issn = {1367-2630},
  doi = {10.1088/1367-2630/11/2/025016},
  urldate = {2025-11-06}
}

@article{guEffectiveModelPairing2025,
  title = {Effective Model and Pairing Tendency in the Bilayer {{Ni-based}} Superconductor \$\textbraceleft\textbackslash mathrm\textbraceleft{{La}}\textbraceright\textbraceright\_\textbraceleft 3\textbraceright\textbraceleft\textbackslash mathrm\textbraceleft{{Ni}}\textbraceright\textbraceright\_\textbraceleft 2\textbraceright\textbraceleft\textbackslash mathrm\textbraceleft{{O}}\textbraceright\textbraceright\_\textbraceleft 7\textbraceright\$},
  author = {Gu, Yuhao and Le, Congcong and Yang, Zhesen and Wu, Xianxin and Hu, Jiangping},
  year = 2025,
  month = may,
  journal = {Physical Review B},
  volume = {111},
  number = {17},
  pages = {174506},
  publisher = {American Physical Society},
  doi = {10.1103/PhysRevB.111.174506},
  urldate = {2025-12-15}
}

@article{houEmergenceHighTemperatureSuperconducting2023,
  title = {Emergence of {{High-Temperature Superconducting Phase}} in {{Pressurized La3Ni2O7 Crystals}}},
  author = {Hou, Jun and Yang, Peng-Tao and Liu, Zi-Yi and Li, Jing-Yuan and Shan, Peng-Fei and Ma, Liang and Wang, Gang and Wang, Ning-Ning and Guo, Hai-Zhong and Sun, Jian-Ping and Uwatoko, Yoshiya and Wang, Meng and Zhang, Guang-Ming and Wang, Bo-Sen and Cheng, Jin-Guang},
  year = 2023,
  month = oct,
  journal = {Chinese Physics Letters},
  volume = {40},
  number = {11},
  pages = {117302},
  publisher = {{Chinese Physical Society and IOP Publishing Ltd}},
  issn = {0256-307X},
  doi = {10.1088/0256-307X/40/11/117302},
  urldate = {2025-12-15},
  langid = {english}
}

@article{jiangHighTemperatureSuperconductivity2024,
  title = {High {{Temperature Superconductivity}} in {{La}}\$\_3\${{Ni}}\$\_2\${{O}}\$\_7\$},
  author = {Jiang, Kun and Wang, Ziqiang and Zhang, Fu-Chun},
  year = 2024,
  month = jan,
  journal = {Chinese Physics Letters},
  volume = {41},
  number = {1},
  eprint = {2308.06771},
  primaryclass = {cond-mat},
  pages = {017402},
  issn = {0256-307X, 1741-3540},
  doi = {10.1088/0256-307X/41/1/017402},
  urldate = {2025-12-15},
  archiveprefix = {arXiv}
}

@article{koSignaturesAmbientPressure2025,
  title = {Signatures of Ambient Pressure Superconductivity in Thin Film {{La3Ni2O7}}},
  author = {Ko, Eun Kyo and Yu, Yijun and Liu, Yidi and Bhatt, Lopa and Li, Jiarui and Thampy, Vivek and Kuo, Cheng-Tai and Wang, Bai Yang and Lee, Yonghun and Lee, Kyuho and Lee, Jun-Sik and Goodge, Berit H. and Muller, David A. and Hwang, Harold Y.},
  year = 2025,
  month = feb,
  journal = {Nature},
  volume = {638},
  number = {8052},
  pages = {935--940},
  issn = {0028-0836, 1476-4687},
  doi = {10.1038/s41586-024-08525-3},
  urldate = {2025-11-03},
  langid = {english}
}

@article{lechermannElectronicCorrelationsSuperconducting2023,
  title = {Electronic Correlations and Superconducting Instability in \$\textbraceleft\textbackslash mathrm\textbraceleft{{La}}\textbraceright\textbraceright\_\textbraceleft 3\textbraceright\textbraceleft\textbackslash mathrm\textbraceleft{{Ni}}\textbraceright\textbraceright\_\textbraceleft 2\textbraceright\textbraceleft\textbackslash mathrm\textbraceleft{{O}}\textbraceright\textbraceright\_\textbraceleft 7\textbraceright\$ under High Pressure},
  author = {Lechermann, Frank and Gondolf, Jannik and B{\"o}tzel, Steffen and Eremin, Ilya M.},
  year = 2023,
  month = nov,
  journal = {Physical Review B},
  volume = {108},
  number = {20},
  pages = {L201121},
  publisher = {American Physical Society},
  doi = {10.1103/PhysRevB.108.L201121},
  urldate = {2025-12-15}
}

@article{leeInfinitelayer$mathrmLamathrmNimathrmO_2$$mathrmNi^1+$2004,
  title = {Infinite-Layer \$\textbackslash mathrm\textbraceleft{{La}}\textbraceright\textbackslash mathrm\textbraceleft{{Ni}}\textbraceright\textbraceleft\textbackslash mathrm\textbraceleft{{O}}\textbraceright\textbraceright\_\textbraceleft 2\textbraceright\$: \$\textbraceleft\textbackslash mathrm\textbraceleft{{Ni}}\textbraceright\textbraceright\textasciicircum\textbraceleft 1+\textbraceright\$ Is Not \$\textbraceleft\textbackslash mathrm\textbraceleft{{Cu}}\textbraceright\textbraceright\textasciicircum\textbraceleft 2+\textbraceright\$},
  shorttitle = {Infinite-Layer \$\textbackslash mathrm\textbraceleft{{La}}\textbraceright\textbackslash mathrm\textbraceleft{{Ni}}\textbraceright\textbraceleft\textbackslash mathrm\textbraceleft{{O}}\textbraceright\textbraceright\_\textbraceleft 2\textbraceright\$},
  author = {Lee, K.-W. and Pickett, W. E.},
  year = 2004,
  month = oct,
  journal = {Physical Review B},
  volume = {70},
  number = {16},
  pages = {165109},
  publisher = {American Physical Society},
  doi = {10.1103/PhysRevB.70.165109},
  urldate = {2025-12-16}
}

@article{lingNeutronDiffractionStudy2000,
  title = {Neutron {{Diffraction Study}} of {{La 3Ni 2O}} 7: {{Structural Relationships Among}} N=1, 2, and 3 {{Phases La}} N+1 {{Ni nO}} 3 N+1},
  shorttitle = {Neutron {{Diffraction Study}} of {{La 3Ni 2O}} 7},
  author = {Ling, Christopher D. and Argyriou, Dimitri N. and Wu, Guoqing and Neumeier, J. J.},
  year = 2000,
  month = jul,
  journal = {Journal of Solid State Chemistry France},
  volume = {152},
  pages = {517--525},
  publisher = {Elsevier},
  issn = {0022-4596},
  doi = {10.1006/jssc.2000.8721},
  urldate = {2025-12-25}
}

@article{liSuperconductivityInfinitelayerNickelate2019,
  title = {Superconductivity in an Infinite-Layer Nickelate},
  author = {Li, Danfeng and Lee, Kyuho and Wang, Bai Yang and Osada, Motoki and Crossley, Samuel and Lee, Hye Ryoung and Cui, Yi and Hikita, Yasuyuki and Hwang, Harold Y.},
  year = 2019,
  month = aug,
  journal = {Nature},
  volume = {572},
  number = {7771},
  pages = {624--627},
  publisher = {Nature Publishing Group},
  issn = {1476-4687},
  doi = {10.1038/s41586-019-1496-5},
  urldate = {2025-12-15},
  copyright = {2019 The Author(s), under exclusive licence to Springer Nature Limited},
  langid = {english}
}

@article{liu$s^ifmmodepmelsetextpmfi$WavePairingDestructive2023,
  title = {\$\textbraceleft s\textbraceright\textasciicircum\textbraceleft\textbackslash ifmmode\textbackslash pm\textbackslash else\textbackslash textpm\textbackslash fi\textbraceleft\textbraceright\textbraceright\$-{{Wave Pairing}} and the {{Destructive Role}} of {{Apical-Oxygen Deficiencies}} in \$\textbraceleft\textbackslash mathrm\textbraceleft{{La}}\textbraceright\textbraceright\_\textbraceleft 3\textbraceright\textbraceleft\textbackslash mathrm\textbraceleft{{Ni}}\textbraceright\textbraceright\_\textbraceleft 2\textbraceright\textbraceleft\textbackslash mathrm\textbraceleft{{O}}\textbraceright\textbraceright\_\textbraceleft 7\textbraceright\$ under {{Pressure}}},
  author = {Liu, Yu-Bo and Mei, Jia-Wei and Ye, Fei and Chen, Wei-Qiang and Yang, Fan},
  year = 2023,
  month = dec,
  journal = {Physical Review Letters},
  volume = {131},
  number = {23},
  pages = {236002},
  publisher = {American Physical Society},
  doi = {10.1103/PhysRevLett.131.236002},
  urldate = {2025-12-15}
}

@article{liuEvidenceChargeSpin2022,
  title = {Evidence for Charge and Spin Density Waves in Single Crystals of {{La3Ni2O7}} and {{La3Ni2O6}}},
  author = {Liu, Zengjia and Sun, Hualei and Huo, Mengwu and Ma, Xiaoyan and Ji, Yi and Yi, Enkui and Li, Lisi and Liu, Hui and Yu, Jia and Zhang, Ziyou and Chen, Zhiqiang and Liang, Feixiang and Dong, Hongliang and Guo, Hanjie and Zhong, Dingyong and Shen, Bing and Li, Shiliang and Wang, Meng},
  year = 2022,
  month = nov,
  journal = {Science China Physics, Mechanics \& Astronomy},
  volume = {66},
  number = {1},
  pages = {217411},
  issn = {1869-1927},
  doi = {10.1007/s11433-022-1962-4},
  urldate = {2025-12-15},
  langid = {english}
}

@article{luInterlayerCouplingDrivenHighTemperatureSuperconductivity2024,
  title = {Interlayer-{{Coupling-Driven High-Temperature Superconductivity}} in {{La}} 3 {{Ni}} 2 {{O}} 7 under {{Pressure}}},
  author = {Lu, Chen and Pan, Zhiming and Yang, Fan and Wu, Congjun},
  year = 2024,
  month = apr,
  journal = {Physical Review Letters},
  volume = {132},
  number = {14},
  pages = {146002},
  issn = {0031-9007, 1079-7114},
  doi = {10.1103/PhysRevLett.132.146002},
  urldate = {2025-11-06},
  langid = {english}
}

@article{luoBilayerTwoOrbitalModel2023,
  title = {Bilayer {{Two-Orbital Model}} of {{L}} a 3 {{N}} i 2 {{O}} 7 under {{Pressure}}},
  author = {Luo, Zhihui and Hu, Xunwu and Wang, Meng and W{\'u}, W{\'e}i and Yao, Dao-Xin},
  year = 2023,
  month = sep,
  journal = {Physical Review Letters},
  volume = {131},
  number = {12},
  pages = {126001},
  issn = {0031-9007, 1079-7114},
  doi = {10.1103/PhysRevLett.131.126001},
  urldate = {2025-11-03},
  langid = {english}
}

@article{marzariThermalContractionDisordering1999,
  title = {Thermal {{Contraction}} and {{Disordering}} of the {{Al}}(110) {{Surface}}},
  author = {Marzari, Nicola and Vanderbilt, David and De Vita, Alessandro and Payne, M. C.},
  year = 1999,
  month = apr,
  journal = {Physical Review Letters},
  volume = {82},
  number = {16},
  pages = {3296--3299},
  publisher = {American Physical Society},
  doi = {10.1103/PhysRevLett.82.3296},
  urldate = {2025-12-14}
}

@article{mostofiWannier90ToolObtaining2008,
  title = {Wannier90: {{A}} Tool for Obtaining Maximally-Localised {{Wannier}} Functions},
  shorttitle = {Wannier90},
  author = {Mostofi, Arash A. and Yates, Jonathan R. and Lee, Young-Su and Souza, Ivo and Vanderbilt, David and Marzari, Nicola},
  year = 2008,
  month = may,
  journal = {Computer Physics Communications},
  volume = {178},
  number = {9},
  pages = {685--699},
  issn = {00104655},
  doi = {10.1016/j.cpc.2007.11.016},
  urldate = {2025-11-06},
  copyright = {https://www.elsevier.com/tdm/userlicense/1.0/},
  langid = {english}
}

@article{ohTypeII$tensuremathJ$Model2023,
  title = {Type-{{II}} \$t\textbackslash ensuremath\textbraceleft -\textbraceright{{J}}\$ Model and Shared Superexchange Coupling from {{Hund}}'s Rule in Superconducting \$\textbraceleft\textbackslash mathrm\textbraceleft{{La}}\textbraceright\textbraceright\_\textbraceleft 3\textbraceright\textbraceleft\textbackslash mathrm\textbraceleft{{Ni}}\textbraceright\textbraceright\_\textbraceleft 2\textbraceright\textbraceleft\textbackslash mathrm\textbraceleft{{O}}\textbraceright\textbraceright\_\textbraceleft 7\textbraceright\$},
  author = {Oh, Hanbit and Zhang, Ya-Hui},
  year = 2023,
  month = nov,
  journal = {Physical Review B},
  volume = {108},
  number = {17},
  pages = {174511},
  publisher = {American Physical Society},
  doi = {10.1103/PhysRevB.108.174511},
  urldate = {2025-12-15}
}

@article{osadaStraintuningSuperconductivityLa3Ni2O72025,
  title = {Strain-Tuning for Superconductivity in {{La3Ni2O7}} Thin Films},
  author = {Osada, Motoki and Terakura, Chieko and Kikkawa, Akiko and Nakajima, Masamichi and Chen, Hsiao-Yi and Nomura, Yusuke and Tokura, Yoshinori and Tsukazaki, Atsushi},
  year = 2025,
  month = jun,
  journal = {Communications Physics},
  volume = {8},
  number = {1},
  pages = {251},
  publisher = {Nature Publishing Group},
  issn = {2399-3650},
  doi = {10.1038/s42005-025-02154-6},
  urldate = {2025-11-05},
  copyright = {2025 The Author(s)},
  langid = {english}
}

@article{perdewGeneralizedGradientApproximation1996,
  title = {Generalized {{Gradient Approximation Made Simple}}},
  author = {Perdew, John P. and Burke, Kieron and Ernzerhof, Matthias},
  year = 1996,
  month = oct,
  journal = {Physical Review Letters},
  volume = {77},
  number = {18},
  pages = {3865--3868},
  issn = {0031-9007, 1079-7114},
  doi = {10.1103/PhysRevLett.77.3865},
  urldate = {2025-11-06},
  copyright = {http://link.aps.org/licenses/aps-default-license},
  langid = {english}
}

@article{pizziWannier90CommunityCode2020,
  title = {Wannier90 as a Community Code: New Features and Applications},
  shorttitle = {Wannier90 as a Community Code},
  author = {Pizzi, Giovanni and Vitale, Valerio and Arita, Ryotaro and Bl{\"u}gel, Stefan and Freimuth, Frank and G{\'e}ranton, Guillaume and Gibertini, Marco and Gresch, Dominik and Johnson, Charles and Koretsune, Takashi and {Iba{\~n}ez-Azpiroz}, Julen and Lee, Hyungjun and Lihm, Jae-Mo and Marchand, Daniel and Marrazzo, Antimo and Mokrousov, Yuriy and Mustafa, Jamal I and Nohara, Yoshiro and Nomura, Yusuke and Paulatto, Lorenzo and Ponc{\'e}, Samuel and Ponweiser, Thomas and Qiao, Junfeng and Th{\"o}le, Florian and Tsirkin, Stepan S and Wierzbowska, Ma{\l}gorzata and Marzari, Nicola and Vanderbilt, David and Souza, Ivo and Mostofi, Arash A and Yates, Jonathan R},
  year = 2020,
  month = apr,
  journal = {Journal of Physics: Condensed Matter},
  volume = {32},
  number = {16},
  pages = {165902},
  issn = {0953-8984, 1361-648X},
  doi = {10.1088/1361-648X/ab51ff},
  urldate = {2025-11-03}
}

@article{puphalUnconventionalCrystalStructure2024,
  title = {Unconventional {{Crystal Structure}} of the {{High-Pressure Superconductor La}} 3 {{Ni}} 2 {{O}} 7},
  author = {Puphal, P. and Reiss, P. and Enderlein, N. and Wu, Y.-M. and Khaliullin, G. and Sundaramurthy, V. and Priessnitz, T. and Knauft, M. and Suthar, A. and Richter, L. and Isobe, M. and Van Aken, P. A. and Takagi, H. and Keimer, B. and Suyolcu, Y. E. and Wehinger, B. and Hansmann, P. and Hepting, M.},
  year = 2024,
  month = oct,
  journal = {Physical Review Letters},
  volume = {133},
  number = {14},
  pages = {146002},
  issn = {0031-9007, 1079-7114},
  doi = {10.1103/PhysRevLett.133.146002},
  urldate = {2025-11-03},
  langid = {english}
}

@article{quBilayer$ttextensuremathJtextensuremathJ_ensuremathperp$Model2024,
  title = {Bilayer \$\textbraceleft t\textbackslash text\textbraceleft\textbackslash ensuremath\textbraceleft -\textbraceright\textbraceright{{J}}\textbackslash text\textbraceleft\textbackslash ensuremath\textbraceleft -\textbraceright\textbraceright{{J}}\textbraceright\_\textbraceleft\textbackslash ensuremath\textbraceleft\textbackslash perp\textbraceright\textbraceright\$ {{Model}} and {{Magnetically Mediated Pairing}} in the {{Pressurized Nickelate}} \$\textbraceleft\textbackslash mathrm\textbraceleft{{La}}\textbraceright\textbraceright\_\textbraceleft 3\textbraceright\textbraceleft\textbackslash mathrm\textbraceleft{{Ni}}\textbraceright\textbraceright\_\textbraceleft 2\textbraceright\textbraceleft\textbackslash mathrm\textbraceleft{{O}}\textbraceright\textbraceright\_\textbraceleft 7\textbraceright\$},
  author = {Qu, Xing-Zhou and Qu, Dai-Wei and Chen, Jialin and Wu, Congjun and Yang, Fan and Li, Wei and Su, Gang},
  year = 2024,
  month = jan,
  journal = {Physical Review Letters},
  volume = {132},
  number = {3},
  pages = {036502},
  publisher = {American Physical Society},
  doi = {10.1103/PhysRevLett.132.036502},
  urldate = {2025-12-15}
}

@article{rhodesStructuralRoutesStabilize2024,
  title = {Structural Routes to Stabilize Superconducting \$\textbraceleft\textbackslash mathrm\textbraceleft{{La}}\textbraceright\textbraceright\_\textbraceleft 3\textbraceright\textbraceleft\textbackslash mathrm\textbraceleft{{Ni}}\textbraceright\textbraceright\_\textbraceleft 2\textbraceright\textbraceleft\textbackslash mathrm\textbraceleft{{O}}\textbraceright\textbraceright\_\textbraceleft 7\textbraceright\$ at Ambient Pressure},
  author = {Rhodes, Luke C. and Wahl, Peter},
  year = 2024,
  month = apr,
  journal = {Physical Review Materials},
  volume = {8},
  number = {4},
  pages = {044801},
  publisher = {American Physical Society},
  doi = {10.1103/PhysRevMaterials.8.044801},
  urldate = {2025-12-15}
}

@article{sakakibaraPossibleHigh$T_c$2024,
  title = {Possible {{High}} \$\textbraceleft{{T}}\textbraceright\_\textbraceleft c\textbraceright\$ {{Superconductivity}} in \$\textbraceleft\textbackslash mathrm\textbraceleft{{La}}\textbraceright\textbraceright\_\textbraceleft 3\textbraceright\textbraceleft\textbackslash mathrm\textbraceleft{{Ni}}\textbraceright\textbraceright\_\textbraceleft 2\textbraceright\textbraceleft\textbackslash mathrm\textbraceleft{{O}}\textbraceright\textbraceright\_\textbraceleft 7\textbraceright\$ under {{High Pressure}} through {{Manifestation}} of a {{Nearly Half-Filled Bilayer Hubbard Model}}},
  author = {Sakakibara, Hirofumi and Kitamine, Naoya and Ochi, Masayuki and Kuroki, Kazuhiko},
  year = 2024,
  month = mar,
  journal = {Physical Review Letters},
  volume = {132},
  number = {10},
  pages = {106002},
  publisher = {American Physical Society},
  doi = {10.1103/PhysRevLett.132.106002},
  urldate = {2025-12-15}
}

@article{shaoBandStructurePairing2025,
  title = {Band {{Structure}} and {{Pairing Nature}} of {{La}}\$\_3\${{Ni}}\$\_2\${{O}}\$\_7\$ {{Thin Film}} at {{Ambient Pressure}}},
  author = {Shao, Zhi-Yan and Liu, Yu-Bo and Liu, Min and Yang, Fan},
  year = 2025,
  month = jul,
  journal = {Physical Review B},
  volume = {112},
  number = {2},
  eprint = {2501.10409},
  primaryclass = {cond-mat},
  pages = {024506},
  issn = {2469-9950, 2469-9969},
  doi = {10.1103/9t6n-jqr5},
  urldate = {2025-12-07},
  archiveprefix = {arXiv}
}

@article{sunSignaturesSuperconductivity802023,
  title = {Signatures of Superconductivity near 80 {{K}} in a Nickelate under High Pressure},
  author = {Sun, Hualei and Huo, Mengwu and Hu, Xunwu and Li, Jingyuan and Liu, Zengjia and Han, Yifeng and Tang, Lingyun and Mao, Zhongquan and Yang, Pengtao and Wang, Bosen and Cheng, Jinguang and Yao, Dao-Xin and Zhang, Guang-Ming and Wang, Meng},
  year = 2023,
  month = sep,
  journal = {Nature},
  volume = {621},
  number = {7979},
  pages = {493--498},
  issn = {0028-0836, 1476-4687},
  doi = {10.1038/s41586-023-06408-7},
  urldate = {2025-11-03},
  langid = {english}
}

@misc{vaulxPressureStrainEffects2025,
  title = {Pressure and Strain Effects on the \$\textbackslash textit\textbraceleft ab Initio\textbraceright\$ \${{GW}}\$ Electronic Structure of {{La}}\$\_3\${{Ni}}\$\_2\${{O}}\$\_7\$},
  author = {de Vaulx, Jean-Baptiste and Meier, Quintin N. and Toulemonde, Pierre and Cano, Andr{\'e}s and Olevano, Valerio},
  year = 2025,
  month = sep,
  number = {arXiv:2504.21651},
  eprint = {2504.21651},
  primaryclass = {cond-mat},
  publisher = {arXiv},
  doi = {10.48550/arXiv.2504.21651},
  urldate = {2025-11-29},
  archiveprefix = {arXiv}
}

@article{wangPressureInducedSuperconductivityPolycrystalline2024,
  title = {Pressure-{{Induced Superconductivity In Polycrystalline}} \$\textbraceleft\textbackslash mathrm\textbraceleft{{La}}\textbraceright\textbraceright\_\textbraceleft 3\textbraceright\textbraceleft\textbackslash mathrm\textbraceleft{{Ni}}\textbraceright\textbraceright\_\textbraceleft 2\textbraceright\textbraceleft\textbackslash mathrm\textbraceleft{{O}}\textbraceright\textbraceright\_\textbraceleft 7\textbackslash ensuremath\textbraceleft -\textbraceright\textbackslash ensuremath\textbraceleft\textbackslash delta\textbraceright\textbraceright\$},
  author = {Wang, G. and Wang, N. N. and Shen, X. L. and Hou, J. and Ma, L. and Shi, L. F. and Ren, Z. A. and Gu, Y. D. and Ma, H. M. and Yang, P. T. and Liu, Z. Y. and Guo, H. Z. and Sun, J. P. and Zhang, G. M. and Calder, S. and Yan, J.-Q. and Wang, B. S. and Uwatoko, Y. and Cheng, J.-G.},
  year = 2024,
  month = mar,
  journal = {Physical Review X},
  volume = {14},
  number = {1},
  pages = {011040},
  publisher = {American Physical Society},
  doi = {10.1103/PhysRevX.14.011040},
  urldate = {2025-12-15}
}

@article{wangStructureResponsibleSuperconducting2024,
  title = {Structure {{Responsible}} for the {{Superconducting State}} in {{La}}{\textsubscript{3}} {{Ni}}{\textsubscript{2}} {{O}}{\textsubscript{7}} at {{High-Pressure}} and {{Low-Temperature Conditions}}},
  author = {Wang, Luhong and Li, Yan and Xie, Sheng-Yi and Liu, Fuyang and Sun, Hualei and Huang, Chaoxin and Gao, Yang and Nakagawa, Takeshi and Fu, Boyang and Dong, Bo and Cao, Zhenhui and Yu, Runze and Kawaguchi, Saori I. and Kadobayashi, Hirokazu and Wang, Meng and Jin, Changqing and Mao, Ho-kwang and Liu, Haozhe},
  year = 2024,
  month = mar,
  journal = {Journal of the American Chemical Society},
  volume = {146},
  number = {11},
  pages = {7506--7514},
  issn = {0002-7863, 1520-5126},
  doi = {10.1021/jacs.3c13094},
  urldate = {2025-12-07},
  copyright = {https://doi.org/10.15223/policy-029},
  langid = {english}
}

@article{xiaSensitiveDependencePairing2025,
  title = {Sensitive Dependence of Pairing Symmetry on {{Ni-eg}} Crystal Field Splitting in the Nickelate Superconductor {{La3Ni2O7}}},
  author = {Xia, Chengliang and Liu, Hongquan and Zhou, Shengjie and Chen, Hanghui},
  year = 2025,
  month = jan,
  journal = {Nature Communications},
  volume = {16},
  number = {1},
  pages = {1054},
  issn = {2041-1723},
  doi = {10.1038/s41467-025-56206-0},
  urldate = {2025-12-03},
  langid = {english}
}

@article{xieStrongInterlayerMagnetic2024,
  title = {Strong Interlayer Magnetic Exchange Coupling in {{La3Ni2O7-$\delta$}} Revealed by Inelastic Neutron Scattering},
  author = {Xie, Tao and Huo, Mengwu and Ni, Xiaosheng and Shen, Feiran and Huang, Xing and Sun, Hualei and Walker, Helen C. and Adroja, Devashibhai and Yu, Dehong and Shen, Bing and He, Lunhua and Cao, Kun and Wang, Meng},
  year = 2024,
  month = oct,
  journal = {Science Bulletin},
  volume = {69},
  number = {20},
  pages = {3221--3227},
  issn = {2095-9281},
  doi = {10.1016/j.scib.2024.07.030},
  langid = {english},
  pmid = {39174404}
}

\end{document}